\documentclass[a4paper,12pt]{article}

\usepackage{amssymb,latexsym}

\makeatletter \@addtoreset{equation}{section} \makeatother
\renewcommand{\theequation}{\thesection.\arabic{equation}}

\newcounter{parentequation}

\begin{document}

\begin{titlepage}

\thispagestyle{empty}

\begin{flushright}
\hfill{CERN-PH-TH/2005-149} \\
\hfill{hep-th/0509052}
\end{flushright}

\vspace{35pt}

\begin{center}{ \LARGE{\bf 
Scherk--Schwarz reduction of M-theory \\[4mm]
on G$_{2}$-manifolds with fluxes
}}

\vspace{60pt}

{\bf 
Gianguido Dall'Agata$^{\ddagger}$ 
\ and \ Nikolaos 
Prezas$^{\star}$}

\vspace{30pt}

{\it  $\ddagger$
Physics Department,\\
Theory Unit, CERN, \\
CH 1211, Geneva 23, \\
Switzerland}

\vspace{15pt}

{\it  $\star$
Institut de Physique,\\
Universit\'e de Neuch\^atel, \\
Neuch\^atel, CH-2000, \\
Switzerland }

\vspace{40pt}

{ABSTRACT}

\end{center}

\vspace{10pt}

We analyse the 4-dimensional effective supergravity theories obtained
from the Scherk--Schwarz reduction of M-theory on twisted 7-tori in
the presence of 4-form fluxes.
We implement the appropriate orbifold projection that preserves a
G$_2$-structure on the internal 7-manifold and truncates the effective
field theory to an ${\cal N} = 1, \; D=4$ supergravity.
We provide a detailed account of the effective supergravity with
explicit expressions for the K\"ahler potential and the superpotential
in terms of the fluxes and of the geometrical data of the internal
manifold.
Subsequently, we explore the landscape of vacua of M-theory
compactifications on twisted tori, where we emphasize the role of
geometric fluxes and discuss the validity of the bottom-up approach.
Finally, by reducing along isometries of the internal 7-manifold, we
obtain superpotentials for the corresponding type IIA backgrounds.

\end{titlepage}

\newpage

\baselineskip 6 mm

\section{Introduction}

In the ordinary approach to 4-dimensional effective theories coming
from flux compactifications, the effect of the fluxes is described by
a mass deformation of the moduli fields.
The moduli one considers are those of the special holonomy manifolds
specifying the background at zero flux.
This procedure obviously neglects the backreaction of the fluxes on
the geometry and is based on the assumption that, for sufficiently
``small'' fluxes, the original moduli can still be used as the
deformation degrees of freedom for the new backgrounds.
This assumption is well justified in some cases, for example for type
IIB compactifications on Calabi--Yau manifolds, where the only effect
of the introduction of non-trivial 3-form fluxes is a conformal
rescaling of the Calabi--Yau metric.
There are, however, several examples of interesting geometries and
models which cannot be described in this way, because the introduction
of fluxes implies a drastic change in the topology of the underlying
manifold and consequently of the moduli fields.

As of today we have exhaustive classifications of the differential and
topological properties of the geometries that preserve some
supersymmetry in the presence of fluxes.
Unfortunately, very few explicit examples of manifolds satisfying the
corresponding
differential and topological constraints have been
found so far.
In this respect, twisted tori provide a simple, but already quite
non--trivial, example of such geometries.
From the point of view of connecting the properties of the effective
theories to the geometry of the internal manifold they are even more
valuable, because of their relation to
Scherk--Schwarz reductions
\cite{Scherk:1978ta,Scherk:1979zr,Kaloper:1999yr}.
In the context of flux compactifications twisted tori were 
revived in \cite{Kachru:2002sk,Derendinger:2004jn}.

Flux compactifications on twisted tori are described by gauged
supergravity models in four dimensions, where the moduli fields couple
and receive masses from both the non-trivial connection of the
internal manifold and the form fluxes.
This fact has been crucial to obtain a simple type IIA model where all
the bulk moduli are stabilized by using gauged supergravity techniques
\cite{Villadoro:2005cu}.
Although it is not yet clear whether this is really sufficient to
stabilize all the moduli of the effective theory, there are by now
several examples where this result is obtained by incorporating
non-perturbative effects \cite{DeWolfe:2005uu,Aspinwall:2005ad}, or 
by using a coset compactification of massive type IIA theory 
\cite{House:2005yc}.
These examples show precisely the relevance of using the appropriate
geometries for flux compactifications, as the introduction of fluxes
is compensated by a twist in the geometry that changes the topological
structure of the tori.

The moduli stabilization problem as well as any attempt to obtain some
vacuum statistics depend heavily on the form of the (super)-potential
of the effective theory.
These potentials have a very simple form when Calabi--Yau or G$_2$
compactifications of string or M-theory are considered
\cite{Gukov:1999gr,Gukov:1999ya,Acharya:2000ps,Beasley:2002db}, while
they become considerably more involved when the internal manifold is
deformed away from special holonomy
\cite{Gurrieri:2002wz,Cardoso:2002hd,Becker:2003gq,Cardoso:2003af}.
In these cases, the derivation of the (super)-potential is less
clear-cut since it is based on several assumptions concerning the
topology of the internal space.
However, as suggested in \cite{DallAgata:2005ff}, twisted tori
compactifications can be used to derive such potentials in terms of
the geometric structures of the internal manifold for more general
cases.
The type IIA case was analysed in detail in \cite{Villadoro:2005cu},
providing an explicit form for the potential, which has indeed more
general application \cite{Grana:2005sn}.
Also, the possible group structures allowed for twisted tori
compactifications in the type IIA case, as well as a more detailed
analysis of the vacua and the corresponding moduli stabilization, were
given in \cite{Camara:2005dc}.

So far, we have a very good understanding of the K\"ahler potential
describing the geometry of the moduli space as well as of the
superpotential for type II and heterotic ${\cal N}=1$
compactifications.
Many insights on the corresponding M-theory objects have also been
obtained \cite{Beasley:2002db,House:2004pm,Lambert:2005sh, Lukas:2003dn}, but, as
mentioned above, when the internal geometries are not given by special
holonomy manifolds, the derivations are either incomplete or less
rigorous.

For these reasons we focus here on ${\cal N}=1$ compactifications of
M-theory on twisted tori.
The starting point is the ${\mathbb T}^{7}$ orbifolds of
G$_2$-holonomy constructed by Joyce \cite{Joyce:1996ig,Joyce:1996rm}
and some smooth generalizations thereof\footnote{The physics of the
non-compact version of these backgrounds and other orbifold
compactifications of M-theory was studied in \cite{Acharya:1998pm}.
For the compact models there exists an interesting type IIA
orientifold interpretation \cite{Kachru:2001je}.}.
These models have 7 main moduli but once fluxes are turned on, be they
of the 4-form field-strength or geometrical (i.e.~Scherk--Schwarz
deformations), the deformed backgrounds have less moduli, as some of
the 7 light fields become massive.
These deformed backgrounds, twisted ${\mathbb T}^{7}$, have no longer
G$_2$-holonomy but rather G$_2$-structure.
This implies that the effective theory still preserves ${\cal N} = 1$
supersymmetry at the Lagrangian level.

We discuss the possible group structures that can be obtained by these
geometric deformations and then determine which ones preserve some
supersymmetry.
For this purpose, we derive the form of the K\"ahler potential and
superpotential of the effective theory and discuss its vacua, putting
emphasis on the moduli stabilization problem.
The K\"ahler potential and superpotential are expressed in a simple
way in terms of the complexified G$_2$-structure $C+i\Phi$, where $C$
is the 3-form gauge potential, and the 4-form background flux $g$:
\begin{equation}
e^{-K/3} = \frac17 \int_{X_7} \Phi \wedge \star \Phi , \qquad {\cal W}
= \frac14 \int_{X_7} \left(C + i \Phi\right)\wedge\left[g+ \frac12
d\left(C + i \Phi\right)\right].
\label{summary}
\end{equation}
We will give this derivation in two different ways, providing also an
extension of the pseudo-action description of \cite{Bergshoeff:2001pv}
to M-theory.

The vacua analysis shows that supersymmetric critical points can be
obtained only by using G$_2$-holonomy manifolds for Minkowski vacua or
weak G$_2$-holonomy manifolds for AdS$_4$ vacua.
As we will see, the twisted tori orbifolds we analyse here can only
give the first type of supersymmetric vacua, while non-supersymmetric
ones can be obtained by using different group structures.

A by-product of our study, using the so-called flat groups
as playground, is the clarification of some aspects of
Scherk-Schwarz reductions related to the precise definition
of the moduli fields. In particular, we emphasize the difference
between ordinary Kaluza--Klein compactification on a twisted
torus and Scherk--Schwarz reduction.

Another important by-product of our analysis is the specification of
the conditions that distinguish non-trivial vacua, i.e.~vacua that
correspond to distinct non-trivial backgrounds of the original
11-dimensional theory.
We will actually see that most of the vacua that are obtained by a
simple analysis of the effective theory represent the same
compactification manifold using trivial cohomological transformations.

Finally, we consider the type IIA reduction of our backgrounds
employing the construction of dynamic and fibered G$_2$-structures
from 6-dimensional SU(3)-structure manifolds \cite{Chiossi:2002kf}.
We show that the general expression for the type IIA superpotential
found in the literature can be indeed recovered this way.
Furthermore, we perform the reduction of the twisted ${\mathbb T}^{7}$
superpotentials and compare it with those of the corresponding type
IIA orientifold.
The latter were constructed in \cite{Derendinger:2004jn} using
gauged supergravity methods.
This reduction shows that the M-theory superpotential has generically
more quadratic couplings than the type IIA one.
We discuss a possible geometrical origin of this discrepancy.

The layout of this paper is as follows.
In section 2 we present the twisted 7-tori that will be the focal
point of this work.
In section 3 we find the 4-dimensional potential that describes the
Scherk--Schwarz reduction of M-theory on the 7-torus including fluxes
and we construct the corresponding superpotential.
In section 5 we examine the vacuum structure of our (super)-potentials
and address the issue of moduli stabilization for these backgrounds.
In section 6 we perform the type IIA reduction of our results and
compare it with those of \cite{Derendinger:2004jn}.
Section 7 summarizes our findings and discusses potential future
directions.
Finally, in an appendix, we show how the pseudo-action that we use to
derive the scalar potential of the effective supergravity theory can
be utilized to describe the general gauged supergravity algebras of
Scherk--Schwarz M-theory compactifications on ${\mathbb T}^{7}$ in
terms of the ``dual'' degrees of freedom, avoiding any recourse to
Free Differential Algebras that appear when massive tensor fields
survive in the effective theory.

\section{The model: twisted 7-torus}
\label{secsetup}

In this section we present our model and discuss some of its
geometrical features that will be relevant for our analysis.
The basic idea is to twist the toroidal orbifolds of
\cite{Joyce:1996ig,Joyce:1996rm} away from G$_2$-holonomy, therefore
obtaining a 7-manifold with G$_2$-structure.
Consistency of the orbifold action with the twisting eliminates some
of the possible twists and demands that the G$_2$-structure is
cocalibrated, i.e.~has a coclosed associative 3-form.
The moduli are introduced in analogy with the toroidal case and
according to the Scherk-Schwarz ansatz \cite{Scherk:1979zr,
Kaloper:1999yr}.

\subsection{${\mathbb T}^{7}$ orbifolds with G$_2$-holonomy}

Our starting point is the compact G$_2$-holonomy\footnote{To be
precise, these orbifolds have a discrete holonomy group ${\mathbb
Z}_2\times {\mathbb Z}_2\times {\mathbb Z}_2 \subset G_2$.
Only the manifolds obtained after blowing up the singularities have
G$_2$-holonomy.} manifolds obtained as toroidal orbifolds of the form
$X_7 = {\mathbb T}^7/({\mathbb Z}_2 \times {\mathbb Z}_2 \times
{\mathbb Z}_2)$ \cite{Joyce:1996ig,Joyce:1996rm}.
Let us take $y^I, \;I=5,\ldots,11$ as coordinates on ${\mathbb T}^7$.
Then, the $({\mathbb Z}_2)^3$ action is
\begin{equation}
\begin{array}{rcl}
{\mathbb Z}_2 (y^I) &=& \{-y^5,-y^6,-y^7,-y^8,c+y^9,y^{10},y^{11}\},\\[3mm]
{\mathbb Z}^\prime_2 (y^I) &=& 
\{a_1-y^5,a_2-y^6,y^7,y^8,c-y^9,c-y^{10},c + y^{11}\},\\[3mm]
{\mathbb Z}^{\prime\prime}_2 (y^I) &=& 
\{a_3-y^5,y^6,a_4-y^7,y^8,a_5-y^9,c+y^{10},-y^{11}\},
\end{array}
\label{orbifold}
\end{equation}
where the coefficients $a_i$ and $c$ can be either $0$ or $1/2$.
This orbifold has singularities that can be either blown-up or
eliminated by turning on $c$ and $a_4$ simultaneously.
Then the orbifold action becomes free and produces a smooth 7-manifold.
It is easy to show that the three $\mathbb{Z}_2$ groups square to the
identity and commute.
The (untwisted) Betti numbers of $X_7$ are independent of $a_i$ and
$c$; they are given by $b_1(X_7) = 0$, $b_2(X_7) = 0$, $b_3(X_7) = 7$.
Notice that resolving the singularities yields extra harmonic forms
and the Betti numbers change.
Here, we will consider only the bulk (i.e.~untwisted) cohomologies and
the corresponding geometrical and physical moduli.

The orbifold action (\ref{orbifold}) preserves a G$_2$-structure,
given by the G$_2$ invariant combination of the 7 surviving 3-forms.
In terms of vielbeins $e^I=R_I dy^I,\;I=5,\ldots,11$, the associated 3-form
$\Phi$ reads
\begin{equation}
\Phi = e^{5}\wedge e^{6}\wedge e^{11}-e^{7}\wedge e^{8}\wedge 
e^{11}-e^{9}\wedge e^{10}\wedge e^{11}+e^{5}\wedge e^{7}\wedge 
e^{10}-e^{6}\wedge e^{7}\wedge e^{9}+e^{5}\wedge e^{8}\wedge 
e^{9}+e^{6}\wedge e^{8}\wedge e^{10}.
\label{Phi}
\end{equation}
This form is obviously closed and coclosed, which implies that the
holonomy group of $X_7$ is contained in G$_2$.
Actually, because of the orbifold projections, this is the only 3-form
(up to signs) that is G$_2$-invariant, and therefore the holonomy
group is exactly G$_2$.
Consequently, the effective 4-dimensional theory obtained by
compactification on $X_7$ can be recast in the form of an ${\cal N}=
1$ supergravity coupled to matter.
This can also be verified by computing the spectrum of the surviving
fields.

Using the information from the Betti numbers, we deduce that the
effective theory for the bulk fields contains only the ${\cal N} = 1$ graviton multiplet
and 7 chiral multiplets $T_i$.
This follows from the generic decomposition of the M-theory 3-form:
\begin{equation}
C  = A^{\alpha}(x) \wedge \omega_{\alpha}(y)+\tau_i(x) \phi^i(y),
\label{decomp}
\end{equation}
where $\omega_{\alpha}$ and 
$\phi_{i}$ span $H^{2}(X_7)$ and 
$H^{3}(X_7)$ respectively,
and from the surviving metric components:
\begin{equation}
g_{\mu\nu}, \quad g_{II}.
\label{metrdeco}
\end{equation}

Since $b_2(X_7)=0$ there are no vector multiplets.
The chiral multiplets contain the moduli of the theory parametrizing
the coset $\left[\frac{SU(1,1)}{U(1)}\right]^{7}$.
The imaginary parts $\tau_i$ of these moduli come from the reduction
(\ref{decomp}) of the M-theory 3-form potential on the internal
manifold:
\begin{equation}
\begin{array}{rcl}
C = &-&\tau_1 dy^5 \wedge dy^6 \wedge dy^{11}+\tau_2 dy^7 \wedge dy^8 
\wedge dy^{11}+\tau_3 dy^9 \wedge dy^{10} \wedge dy^{11}+\tau_4 dy^6 
\wedge dy^7 \wedge dy^9\\[3mm]
&-&\tau_5 dy^6 \wedge dy^8 \wedge 
dy^{10}-\tau_6 dy^5 \wedge dy^7 \wedge dy^{10}-\tau_7 dy^5 \wedge 
dy^8 \wedge dy^9.
\end{array}
\label{Cform}
\end{equation}
The real parts $t_i$ are associated with the internal metric components
and parametrize the volumes of the seven surviving 3-cycles:
\begin{equation}
\Phi  = t_i(x) \phi^i(y).
\end{equation}
The holomorphic combinations $T_i = t_i + i \, \tau_i$ can be thought
of as expansion parameters of a ``complexified'' G$_2$-form
\begin{equation}
C + i \,\Phi = i \, T_i(x) 
\,\phi^i(y).
\end{equation}
This is analogous to the complexification of K\"ahler moduli in type
II string theories.

In terms of the radii $R_I$ of the 7 circles composing ${\mathbb 
T}^7$,
we have $e^I=R_I dy^I$ and hence
\begin{eqnarray}
|t_1| &=& R_5 R_6 R_{11}, \;\;\; |t_2|=R_7 R_8 R_{11}, \;\;\; |t_3| = 
R_9 R_{10} R_{11},\;\;\;
|t_4|=R_6 R_7 R_9,\nonumber\\
|t_5|&=&R_6 R_8 R_{10}, \;\;\; |t_6|=R_5 
R_7 R_{10},\;\;\;
|t_7|=R_5 R_8 R_9.
\end{eqnarray}
We can also write the vielbeins in terms of the moduli as
\begin{equation}
\begin{array}{rclcrclcrcl}
e^{5} &=& \displaystyle \sqrt{\frac{t_1 t_6 t_7}{V}}dy^5, &\quad & e^{6} &=& \displaystyle 
\sqrt{\frac{t_1 t_4 t_5}{V}}dy^6 ,&\quad& e^{7} &=&\displaystyle  
\sqrt{\frac{t_2 t_4 t_6}{V}}dy^7,\\[4mm]
e^{8} &=& \displaystyle \sqrt{\frac{t_2 t_5 t_7}{V}}dy^8 , &\quad & e^{9} &=& \displaystyle 
\sqrt{\frac{t_3 t_4 t_7}{V}}dy^9,&\quad&  e^{10} &=& \displaystyle 
\sqrt{\frac{t_3 t_5 t_6}{V}}dy^{10} ,\\[4mm]
e^{11} &=& \displaystyle \sqrt{\frac{t_1 t_2 t_3}{V}}dy^{11}, &\quad& 
V &=& (t_1 \cdots t_7)^{1/3}.
\end{array}
\label{viel}
\end{equation}

We need to impose the conditions $t_i \neq 0$, $\forall i$, so as to
avoid metric degeneracies.
Furthermore, the combinations under the square roots in (\ref{viel})
must be positive (which means that either all the $t_i$ are positive
or four of them are negative and two positive).

\subsection{Twisting and G$_2$-structures}

We now consider the orbifold action (\ref{orbifold}) on a twisted
$\mathbb{T}^7$.
Physically, the twisting can be interpreted as performing a
Scherk--Schwarz reduction on the {\em original} untwisted torus.
Practically, it simply means that we have to replace the straight
differentials $dy^I$ in (\ref{viel}) by twisted ones, i.e.~1-forms 
$\eta^I$ that satisfy
\begin{equation}
d \eta^{I} - \frac12 \tau_{JK}^{I} \eta^J \wedge \eta^K = 0
\label{torsion}
\end{equation}
with structure-constant parameters $\tau_{JK}^{I}=-\tau_{KJ}^I$
\cite{Kaloper:1999yr,DallAgata:2005ff,Hull:2005hk,Villadoro:2005cu}.

For a generic Scherk--Schwarz reduction, these constants are
constrained by
\begin{equation}
\tau_{IJ}^{I} = 0, \qquad  \tau_{[IJ}^{L}\tau_{K]L}^{M}= 0.
\label{const}
\end{equation}
The first condition guarantees that we can integrate out the
dependence on the coordinates of the internal manifold and obtain a
Lagrangian 4-dimensional theory.
This condition can be relaxed when we consider a reduction of the
equations of motion without requiring a well-defined Lagrangian.
The second condition stems from the nilpotency of exterior
differentiation and therefore  it is relevant 
when we want to interpret our Scherk--Schwarz
reduction as a compactification on a twisted torus.
Furthermore, the consistency of (\ref{const}) with the orbifold action
implies that we can keep only those structure constants
$\tau_{JK}^{I}$ with indices $(I,J,K)$ acted upon as $(+,+,+),
(+,-,-),(-,-,+),(-,+,-)$ under (\ref{orbifold}).
Hence, the only surviving structure constants are
\begin{eqnarray}
& \Big\{ \tau_{510}^{7} , \tau_{710}^{5},\tau_{57}^{10}\Big\},\; 
\Big\{\tau_{79}^{6}, \tau_{69}^{7}, \tau_{67}^{9}\Big\},\;
\Big\{\tau_{59}^{8} , \tau_{58}^{9}, \tau_{89}^{5}\Big\} ,\;
\Big\{\tau_{68}^{10},
\tau_{610}^{8}, \tau_{810}^{6}\Big\}, & \\
& \Big\{\tau_{56}^{11} , 
\tau_{78}^{11}, \tau_{910}^{11}\Big\}, &\\
&
\Big\{\tau_{611}^5, \tau_{511}^{6} ,
\tau_{811}^{7} , \tau_{711}^8, 
\tau_{1011}^{9},  \tau_{911}^{10}\Big\},  &
\label{tausurv}
\end{eqnarray}
Notice that the first condition in (\ref{const}) is automatically
satisfied by the surviving structure constants.

When the twist in the metric is introduced, the space $X_7$ is no
longer a G$_2$-holonomy manifold, but its tangent bundle rather shows
a G$_2$-structure.
This means that the rank-3 invariant form in (\ref{Phi}), written now
in terms of the twisted vielbeins $e^I=R_I \; \eta^I$, is preserved by
the orbifold action and is a singlet of the group structure of the
tangent bundle.
However, a priori $\Phi$ is no longer closed and coclosed; the
deviation from special holonomy is parametrized by 4 intrinsic torsion
classes $W_1, W_7, W_{14}, W_{27}$, which are defined as
\begin{eqnarray}
d\Phi & = & W_1 \, \star\!\Phi - \Phi \wedge W_{7} + W_{27}, 
\label{dphi}\\
d\star\!\Phi & = & \frac43 \star\!\Phi \wedge W_7 + W_{14}, \label{dstarphi}
\end{eqnarray}
and where $\Phi \wedge W_{27} = 0$ and $\Phi \;\lrcorner \;W_{27}=0$.
The subscripts refer to the representations of the various components
under G$_2$.
Note that the Betti numbers of the twisted tori depend on the specific
set of structure constants we choose to turn on and in general are
different than those of the original untwisted $\mathbb{T}^7$.

It is interesting to notice that not all of these torsion classes can
be realized using twisted tori.
By direct inspection of the exterior differentials of $\Phi$ and its
dual, one can actually find that the coassociative form defined on
$X_7$ is closed for the twists (\ref{tausurv}) that survive the
orbifold projection:
\begin{equation}
W_{7} = W_{14} = 0.
\label{classeszero}
\end{equation}
Hence, the allowed G$_2$ structures are cocalibrated \cite{friedrich}, i.e.~they 
satisfy
\begin{eqnarray}
d\Phi & = & W_1 \, \star\!\Phi + W_{27}, 
\label{dphi2}\\
d\star\!\Phi & = & 0. \label{dstarphi2}
\end{eqnarray}
$W_1$ and $W_{27}$ can be explicitly computed in terms of the
geometric fluxes coefficients.
For example (setting all the radii to 1)
\begin{equation}
\begin{array}{rcl}
W_1 &=& \displaystyle \frac27
\left(\tau_{611}^{5}-\tau_{59}^{8}-\tau_{610}^{8}-\tau_{1011}^{9}
-\tau_{67}^{9} +
\tau_{711}^{8}+\tau_{58}^{9}+\tau_{911}^{10}+\tau_{57}^{10}+\tau_{68}^{10}
-\tau_{78}^{11}\right.\\[7mm] 
&&\left. \displaystyle+\tau_{710}^5+\tau_{56}^{11}-\tau_{910}^{11}
+\tau_{89}^{5}-\tau_{511}^{6}-\tau_{79}^6+\tau_{810}^{6}
-\tau_{811}^{7}-\tau_{510}^{7}+\tau_{69}^{7}\right),
\end{array}
\label{W1}
\end{equation}
while the 4-form $W_{27}$ can be also easily obtained from (\ref{dphi2}).
Obviously, not all of these torsions will allow for supersymmetric
vacua.
We will see later, in the analysis of the vacua of the superpotential,
which examples are concretely realized out of the possible ones
described here.

We should stress here that the manifold under consideration has
precisely a G$_2$-structure and not subgroups thereof, since 1- and
2-forms are projected out by the orbifold.
Furthermore, the fact that the G$_2$-structure we obtain is coclosed
is also a consequence of the orbifold projections.
A generic twisted 7-torus can actually contain all the torsion
components in (\ref{dphi}) and (\ref{dstarphi}).
It is also clear that this more generic situation would allow extended
4-dimensional supersymmetries; the analysis should therefore be
refined by using SU(3) or even SU(2) structures.

\section{Potential and superpotential}
\label{secpotential}

In order to derive the effective potential and superpotential in the
presence of both form and geometrical fluxes, we have to reduce the
action of 11-dimensional supergravity on $X_7$.
For what concerns ${\cal N} = 1$ effective theories, various
derivations of the superpotential have been presented either for
G$_2$-holonomy manifolds with 4-form flux turned on
\cite{Acharya:2000ps,Beasley:2002db,Curio:2002ja,Acharya:2002kv,Grimm:2004ua},
or by using weak-G$_2$-holonomy manifolds
\cite{Lambert:2005sh,House:2004pm}, or by reducing 
the supersymmetry transformations \cite{Behrndt:2003zg,Behrndt:2004mx}.
In the following we will extend these results for twisted tori
compactifications.

The reduction of M-theory in the presence of 4-form fluxes $G_4$ is
subtle, since the Chern--Simons terms are modified in the presence of
topologically non-trivial background flux $g$.
It was shown in \cite{DeWolfe:2005uu} that invariance under large
gauge transformations fixes completely the Chern--Simons terms and the
correct action is
\begin{equation}
\begin{array}{rcl}
S_{11d} &=& \displaystyle \frac12 \int d^{11}x \sqrt{-g} R - 
\frac{1}{4}\int (g + dC)
\wedge \star (g+dC) - \frac{1}{12} \int dC \wedge dC \wedge C \\[4mm]
&& \displaystyle - \frac14 \int dC \wedge g \wedge C - \frac14 \int g 
\wedge g \wedge C,
\end{array}
\label{11dsugra}
\end{equation}
where the standard 4-form combination $G_4 = g + dC$ appears only in
the kinetic term, while the Chern--Simons terms cannot be recast in
terms of only $C$ and $G_4$.
These non-trivial Chern--Simons terms appear only when one considers
fluxes $g$ that are closed but not exact.
Obviously the exact part will still appear with the same coefficient
as $dC$.
Notice that varying this action with respect to the 3-form gauge
tensor $C$ the equation of motion for $G_4$ is obtained, which again
depends only on the expected combination $G_4 = g+ dC$.

The reduction of the Einstein kinetic term was first performed in the seminal
paper by Scherk and Schwarz \cite{Scherk:1979zr}; with our
normalizations, it reads
\begin{equation}
V_E = \frac18 \,\frac{1}{e_7} \, \left(2 \, \tau_{JK}^I \tau_{IL}^{J} g^{KL} + \tau_{JK}^{I} \tau_{MN}^{L} 
g_{IL} g^{JM} g^{KN}\right).
\label{Einstein}
\end{equation}
The volume of the internal space $e_7 = (t_1 \ldots t_7)^{1/3}$
appears in the denominator because of the rescaling of the space--time
metric $g_{\mu\nu} \to e_7 \,g_{\mu\nu}$ required to go to the
Einstein frame in 4-dimensions.

Given the expansion of the 4-form field strength vacuum expectation 
value 
\begin{equation}
g = \frac{1}{4!} g_{IJKL} \eta^{I} \wedge \eta^{J}\wedge 
\eta^{K}\wedge \eta^{L},
\label{4formexp}
\end{equation}
which is the appropriate one for a Scherk-Schwarz reduction with fluxes 
\cite{Kaloper:1999yr},
the 4-form kinetic term yields
\begin{equation}
V_K = \frac{1}{96}\, \frac{1}{e_7} \, \left(g_{IJKL} + 6\,\tau_{[IJ}^{P}C_{KL]P}\right) 
g^{II^\prime}g^{JJ^\prime}g^{KK^\prime}g^{LL^\prime} \left(g_{I^\prime 
J^\prime K^\prime L^\prime} + 6\,\tau_{[I^\prime 
J^\prime}^{P^\prime}C_{K^\prime L^\prime]P^\prime}\right) .
\label{V4form}
\end{equation}
As a consequence of the Bianchi identity, the topological flux $g$
must obey
\begin{equation}
\tau_{[IJ}^{P}g_{JKL]P} = 0.
\label{IB}
\end{equation}
Actually, for our model these conditions are identically satisfied,
thanks to the orbifold projections.

Finally, the Chern--Simons contribution to the scalar potential can be
obtained by integrating the 4-form with external legs, which acts as a
Lagrange multiplier: $G_{\mu\nu\rho\sigma} = \lambda
\epsilon_{\mu\nu\rho\sigma}$.
By doing so one obtains
\begin{equation}
V_{{\rm CS}}=\frac{1}{3! 4!}  \frac{1}{e_7} \left[\frac{1}{4! e_7} \, \epsilon^{IJKLMNP} \left(g_{IJKL} + 3\, \tau_{[IJ}^{Q} 
C_{KL]Q}\right) C_{MNP}\right]^2.
\label{VCS}
\end{equation}
What should be noted in this expression is the factor of 2 difference
with the similar term that comes from the ordinary $G_4$ combination
appearing in $V_K$.
This difference is due to the non-standard form of the Chern--Simons
terms, which guarantees gauge invariance under large gauge
transformations.
This difference actually is crucial in order to express the potential
in terms of the superpotential.
Again, this happens only when considering fluxes $g$, which are closed
but not exact.
Any exact part can be reabsorbed in the potential as a constant
background value for the $C_{IJK}$ scalars and appears with the same
coefficients as the parts involving $\tau C$.
From this fact we can also understand that $G_4= 0$ will not be a
critical point of the potential whenever the non-exact part is
non-vanishing $g \neq 0$ (we will show later that the vacua in
\cite{DAuria:2005dd} are trivial redefinitions of the zero-flux
vacuum).

The derivation performed so far is clearly general and gives a
potential for an ${\cal N}= 8$ effective supergravity\footnote{While
this paper was in preparation, a similar result was also obtained in
\cite{DAuria:2005dd} by integrating the equations of motion of
11-dimensional supergravity.}.
However, thanks to the orbifold projections, the sum of the various
terms can be recast in the standard ${\cal N} = 1$ form without
$D$-terms
\begin{equation}
V = e^{K}\left( g^{i \bar \jmath} D_{i} {\cal W} D_{\bar \jmath} 
\overline {\cal W} - 3 |{\cal W}|^2\right),
\label{Pot}
\end{equation}
where $D_i {\cal W} = \partial_i {\cal W} + \partial_i K\, {\cal W}$.
An explicit calculation shows that this happens using the K\"ahler
potential of the $\left[\frac{SU(1,1)}{U(1)}\right]^{7}$ scalar 
manifold:
\begin{equation}
K = -\sum_{i=1}^{7} \log\frac{T_i + \overline T_i}{2}=-\log\left(t_1
t_2 t_3 t_4 t_5 t_6 t_7\right),
\label{Kaehlerbis}
\end{equation}
and the superpotential
\begin{equation}
\begin{array}{rcl}
4 \,{\cal W} &=& - T_5 T_7 \, \tau_{711}^8 - T_6 T_7 \,{
\tau_{611}^5}+T_4 T_7 \, { \tau_{1011}^{9}} \\[2mm] &+& T_1 T_2\, {
\tau_{910}^{11}} + T_1 T_3\,{ \tau_{78}^{11}} -T_2 T_3\, {
\tau_{56}^{11}} - T_1 T_4\, { \tau_{810}^{6}}+T_2 T_4\, {
\tau_{510}^{7}} - T_3 T_4\,{ \tau_{58}^{9}}\\[2mm] &+& T_1 T_5\,{
\tau_{79}^{6}} +T_2 T_5\,{ \tau_{59}^{8}} -T_3 T_5\, { \tau_{57}^{10}}
+T_4 T_5\, { \tau_{511}^{6}} -T_1 T_6 \, { \tau_{89}^{5}} - T_2 T_6\,
{ \tau_{69}^{7}}\\[2mm] &+&T_4 T_6\, { \tau_{811}^{7}} -T_3 T_6\, {
\tau_{68}^{10}} -T_5 T_6\,{ \tau_{911}^{10}} -T_1 T_7 \, {
\tau_{710}^{5}} + T_2 T_7\,{ \tau_{610}^{8}} +T_3 T_7\, {
\tau_{67}^{9}}\\[2mm] &+& {\rm i } \left( T_1 \,{ g_{78910}} - T_2 \,{
g_{56910}} - T_3 \,{ g_{5678}}- T_4 \,{ g_{581011}} + T_5 \,{
g_{57911}} \right.\\[2mm] &+& \left.
T_6 \,{ g_{68911}} + T_7\, { g_{671011}}\right).
\end{array}
\label{Wexpanded}
\end{equation}
The na\"ive reduction generically produces terms that could not be
recast in the ${\cal N}=1$ form (\ref{Pot}), but which vanish once the
orbifold projections and the constraints (\ref{const}) on the torsions
are taken into account.
In order to really consider only the topological flux contributions in
(\ref{Wexpanded}), one should take into account additional constraints
on the allowed values of the flux components, due to the topology
changes driven by the twistings $\tau_{IJ}^{K}$.
For this reason we will allow generic values and then distinguish the
cohomologically different parts.

The potential obtained by plugging (\ref{Wexpanded}) in (\ref{Pot})
does not contain the possible non-trivial cosmological constant
contributions that are due to expectation values of the 4-form on the
external part.
In our framework, since we would like to have a complete control also
on this value, the cosmological term is best described by the
expectation value of the dual 7-form $\tilde G_7$.
This dual form obviously needs a different 11-dimensional formulation from
(\ref{11dsugra}).
The mechanism we decided to use is the M-theory generalization of the
pseudo-action approach, which allows for a dual formulation of type
IIA supergravity \cite{Bergshoeff:2001pv} and which was also used in a
similar reduction in \cite{Villadoro:2005cu}.
This mechanism requires the formulation of a pseudo-action where the
potentials and the dual curvatures appear.
The first are considered as real degrees of freedom, while the dual
curvatures are just ``empty boxes'', which act as multiplicative
constants.
Then, by varying the pseudo-action with respect to the potentials, one
obtains Bianchi identities for the dual curvatures, which can be
solved in terms of dual potentials.
Finally, these solutions are plugged back in the pseudo-action, giving
a real action to use for deriving the equation of motion of the
potentials corresponding to the curvatures in the pseudo-action.

Trying to adapt this procedure in M-theory, we realize that, because
of the Chern--Simons terms, the expected (dual) action for
11-dimensional supergravity cannot be used if we want to describe all
the non-metric degrees of freedom\footnote{An alternative formulation
that allows for this description is given by the duality symmetric
actions constructed in \cite{Bandos:1997gd,Bandos:2003et}.} in terms
of the standard potential $C$ or of the dual one $A$.
However, if one is interested in a split description of the potential
degrees of freedom and of their curvatures, partly by the 3-form and
partly by the 6-form, this approach is effective, as we will see in a
moment.

The starting point is the pseudo-action
\begin{equation}
\begin{array}{rcl}
S_{11d} &=& \displaystyle \frac12 \int d^{11}x \sqrt{-g} R - 
\frac{1}{4}\int G_4
\wedge \star G_4 -\frac{1}{4} \int \tilde G_7 \wedge \star \tilde G_7  \\[4mm]
&& \displaystyle - \frac{1}{12} \int C \wedge \left(d \tilde G_7+G_4 \wedge 
G_4\right) - \int A \wedge d G_4,
\end{array}
\label{pseudoaction}
\end{equation}
where only some couples $\{C,\tilde G\}$ and $\{A,G\}$ are kept and
$G_4$, $\tilde G_7$ are empty boxes for the moment.
For the case at hand, the orbifold projection leaves little choice,
since the only components of the potentials and fluxes that survive
are
\begin{equation}
\begin{array}{l}
C_{\mu\nu\rho}, \quad C_{IJK} , \quad A_{\mu\nu\rho IJK}, \quad 
A_{\mu\nu IJKL}, \\[2mm]
G_{\mu\nu\rho\sigma} , \quad G_{\mu IJK} , \quad G_{IJKL} , \quad 
\widetilde G_{\mu\nu\rho\sigma IJK} , \quad \widetilde G_{\mu\nu\rho IJKL} , 
\quad \widetilde G_{IJKLMNP}.
\end{array}
\label{survpotflux}
\end{equation}
Since we want to obtain a final action that depends on $C_{IJK}$ and
contains the curvatures $G_{IJKL}$, $G_{\mu IJK}$, $G_{IJKLMNP}$, we
start with (\ref{pseudoaction}), where we set to zero all the terms
containing this potential, as well as all the curvatures dual to the
ones we want to appear in the final action.
This means that we set to zero $G_{\mu\nu\rho\sigma}$, $\tilde
G_{\mu\nu\rho\sigma IJK}$ and $\tilde G_{\mu\nu\rho IJKL}$.
By doing so, (\ref{pseudoaction}) contains only 3 potential terms,
depending on $C_{\mu\nu\rho}$, $A_{\mu\nu IJKL}$ and $A_{\mu\nu\rho
IJK}$.
The variation with respect to these components gives the equations of
motion:
\begin{eqnarray}
\partial_\mu \tilde G_{IJKLMNP} +70 \,G_{\mu [IJK} G_{LMNP]} & = & 0, 
\label{i}\\
\partial_{[\mu} G_{\nu]IJK} & = & 0, \label{ii}\\
\partial_{\mu} G_{IJKL} -6\, \tau_{[IJ}^{M} G_{\mu KL]M} & = & 0, \label{iii}
\end{eqnarray}
which can be interpreted as Bianchi identities for the various
curvatures.

Solving the Bianchi identities (\ref{i})--(\ref{iii}) we obtain the
definitions of these curvatures in terms of the basic degrees of
freedom and background fluxes on the internal manifold (we do not
allow for constant background values of the curvatures with mixed
indices):
\begin{eqnarray}
G_{\mu IJK} &=& \partial_{\mu} C_{IJK},\label{soli} \\[2mm]
G_{IJKL} &=& g_{IJKL} + 6 \tau_{[IJ}^M C_{KL]M}, \label{solii} \\[2mm]
\tilde G_{IJKLMNP} &=& g_{IJKLMNP} -
70 \left(g_{[IJKL} + 3\tau_{[IJ}^{Q} C_{KL|Q|}\right)C_{MNP]}.
\label{soliii}
\end{eqnarray}
The first curvature definition is a trivial solution to (\ref{ii}),
and (\ref{solii}) follows from (\ref{iii}) when plugging the solution
(\ref{soli}) in it.
The last one is a bit more tricky, as the equation (\ref{i}) does not
become a total space--time derivative of a single quantity, even after
plugging the solution (\ref{soli}) and (\ref{solii}) in it.
However, inspection of the orbifold conditions shows that $\tau
\left(C \wedge \partial_{\mu}C\right) = 0$, which implies that one can
rewrite (\ref{i}) as a space--time derivative on a single object.

Plugging these solutions back in the pseudo-action
(\ref{pseudoaction}) results in a real action, which gives a
4-dimensional potential with precisely the various terms computed
above.
In addition, the Chern--Simons term is now added by a constant shift,
which depends on the 7-form flux $g_{IJKLMNP}$.
It is also interesting to notice that from this approach we do not
obtain the 4-form Bianchi identity (\ref{IB}).
This, however, is identically satisfied with our orbifold projections
and therefore does not impose additional constraints.
Also notice the factor of 2 between the constant 4-form flux and the
cohomologically trivial part in (\ref{solii}) and (\ref{soliii}).
These are precisely the same factors which enter in (\ref{V4form}) and
(\ref{VCS}) respectively and for which we have already given a
detailed account.

We emphasize here that this framework is the most appropriate for
studying the effect of M5-branes wrapped on 2-cycles of the internal
manifold in the above setup, since these naturally couple to the
6-form flux $A$.
For the orbifold at hand, however, we can easily see that the
$A_{\mu\nu\rho\sigma IJ}$ component is projected out  since there are no
surviving 2-cycles and hence one cannot introduce this type of
M5-branes.
In more general cases, introducing these wrapped M5-branes 
will affect the constraints on
the structure constants of the effective gauge couplings in
4-dimensions.
More specifically, since the Bianchi identity of the 4-form flux will
now be modified by a source term, this will imply a different closure
of the gauge algebra governing the lower dimensional effective theory.

We are now in a position to write the K\"ahler potential and the
superpotential in terms of basic geometrical quantities, using the
definitions (\ref{Phi}) and (\ref{Cform}) and with the straight
differentials $dy^I$ being substituted by the 1-forms $\eta^I$.
The K\"ahler potential reads
\begin{equation}
K = -3 \log \left(\frac17 \int_{X_7} \Phi \wedge \star \Phi\right),
\label{Kaehler}
\end{equation}
and the superpotential is given by
\begin{equation}
{\cal W} = \frac14 \, \int_{X_7} \tilde G_7 + \frac14 \, \int_{X_7}
\left(C + i \, \Phi\right)\wedge\left[g + \frac12 d\left(C+ i \,
\Phi\right)\right],
\label{Wform}
\end{equation}
where the exterior differentiation on the internal space satisfies
(\ref{torsion}) and gives rise to the terms containing the structure
constants $\tau_{IJ}^{K}$.
The K\"ahler potential (\ref{Kaehler}) coincides with the Hitchin
functional describing the space of stable 3-forms for 7-manifolds.
In analogy to type II compactifications on SU(3)-structure manifolds
\cite{Grana:2005ny}, it would be interesting to prove that this
functional describes the moduli space of generic G$_2$-structure
deformations, as suggested by (\ref{Kaehler}).

Some comments are now in order concerning the general form of the
superpotential.
First of all, note that the flux contributes with terms linear in the
holomorphic coordinates while the torsion is coupled to a quadratic
form of the coordinates.
Owing to the structure of the $C$ form, analogous to $\Phi$, we also
see that only the $W_1$ torsion class computed in (\ref{W1})
contributes to ${\cal W}$.
For the same reason, the 4-form flux contributes only with its singlet
part.
This is expected since invariance of ${\cal W}$ demands that only
singlets of G$_2$ should
appear in it.
In the absence of a warp factor, the vanishing of this superpotential,
which is a necessary condition for supersymmetric Minkowski vacua,
implies that these components vanish, in accordance with the analysis
of the supersymmetry variations
\cite{Kaste:2003zd,DallAgata:2003ir,Behrndt:2003zg}.

Secondly, we believe that such an expression has more general validity
 than the twisted tori compactifications used here for its
derivation.
The superpotential for G$_2$-holonomy manifolds with 4-form flux
turned on was computed in \cite{Beasley:2002db}, while
\cite{House:2004pm} extended this result to the case of G$_2$
structure manifolds.
In the latter work, the potential was computed from explicit
compactification only for manifolds with weak G$_2$ holonomy, i.e.
for manifolds with only $W_1$ non-vanishing.
Our derivation here for the twisted 7-torus extends this result to a
situation where both $W_1$ and $W_{27}$ are different from zero and is a
further consistency check on the proposal of \cite{House:2004pm}.
We should mention that the comparison with \cite{Beasley:2002db} in
the limit of G$_2$ holonomy manifolds shows a different factor; this,
however, comes from the assumption in \cite{Beasley:2002db} that $dC$
still gives a non-vanishing contribution to the potential,
and it can be reconciled with that of  \cite{Beasley:2002db}  
after an integration by parts \cite{Lambert:2005sh}.
Furthermore, although our expression for the superpotential contains
the same terms as those in \cite{House:2004pm}, there is a clear
difference between us and \cite{Beasley:2002db} on one side and
\cite{House:2004pm} on the other.
The difference is in the distribution of the various terms in the real
and imaginary components of ${\cal W}$ and is actually crucial for
deriving supersymmetric vacua.

\section{Vacua and interpretation}

As we stated in the introduction, the main purpose of constructing the
effective theory for M-theory Scherk--Schwarz compactifications with
fluxes is to provide an alternative way of determining supersymmetric
backgrounds and studying the landscape of vacua when the internal
manifolds are given by twisted tori.
Hopefully this can help generalizing the results of
\cite{Acharya:2005ez} when the internal manifolds have non-trivial
intrinsic torsion.
In order to do so, the effective theory should capture all the
properties of consistent 11-dimensional vacua.
A very important consequence of this fact is that we should not expect
critical points of the potential corresponding to Minkowski (or de Sitter)
vacua with non-trivial fluxes. Instead, an AdS vacuum may be in principle
possible.
These expectations are due to the no-go theorem of
\cite{deWit:1987xg,Maldacena:2000mw}.

Under quite general assumptions, the only allowed 
compactifications of M-theory to 4-dimensions are
either Minkowski vacua with $G_4 = g_7 = 0$ or Freund--Rubin
solutions, giving AdS$_4$ with $G_4 = 0$ but $g_7 \neq 0$.
Allowing for the presence of source terms like M-branes, may lead to
more general types of supersymmetric configurations, possibly with
non-trivial 4-form flux. 
Another way of bypassing this no-go theorem is the addition of higher
order derivative terms in the action, for example terms involving
higher powers of the Riemann curvature tensor.
Since in the following we will stick to the setup described in section
\ref{secpotential}, where our starting point is pure 11-dimensional
supergravity without higher order corrections, we should expect that
our results be in accord with the no-go theorem and its implications.

\subsection{General properties}

As a first step for the analysis of the vacua, we discuss some generic
features of the superpotentials of the form (\ref{Wexpanded}).
We use a compact version of (\ref{Wexpanded}), which can be written as
\begin{equation}
{\cal W} = \frac12 M_{ij} T^i T^j + i\, G_i T^i +  g_7,
\label{genW}
\end{equation}
where $M$, $G$ and $g_7$ are all real and are associated to the
geometrical, the 4-form, and the 7-form fluxes respectively.
Moreover, the matrix $M$ is symmetric, with zeros along the diagonal.
In this way we can discuss the critical points without 
reference to a particular configuration of geometric and/or form
fluxes.
A similar expression has also appeared in \cite{House:2004pm} but for
the crucial difference in the factor of $i$ in front of
the 4-form fluxes.

The general form (\ref{genW}) can be easily read from
(\ref{Wexpanded}), with the addition of the 7-form flux, but it is
also useful to derive it directly from (\ref{Wform}).
The latter method has the advantage of producing a compact expression
for $M$ (see also \cite{House:2004pm}) and can be used to formulate
the supersymmetry conditions in terms of the torsion classes in
(\ref{dphi}) and (\ref{dstarphi}).
For this purpose, we use a basis of 3-forms $\phi^{i}$ (which
coincides with the basis of seven harmonic 3-forms of the untwisted
manifold when the structure constants are set to zero) and dual
4-forms $\widetilde \phi^{i}$ satisfying\footnote{The 4-forms are not
directly expanded using the $\star \phi^i$ forms, which depend
explicitly on the metric, but rather in the $\widetilde \phi^i$ basis,
which is generically constructed by taking linear combinations of the
$\star \phi^i$ \cite{House:2004pm}.}
\begin{equation}
\int_{X_7} \phi^i \wedge \widetilde \phi^j = \delta^{ij}.
\label{basis}
\end{equation}
Plugging in (\ref{Wform}) the expansions $C+ i \Phi = i\, T_i
\,\phi^{i}$ and $g_4 = 4 G_i \widetilde\phi^i$ yields
\begin{equation}
4 {\cal W} = \frac12 \, T_i \, T_j \, \int_{X_7} \phi^i \wedge d 
\phi^j  + 4i \, T_i G_j \int_{X_7} \phi^i \wedge \widetilde \phi^j +  
\int_{X_7} \widetilde G_7.
\label{Wprima}
\end{equation}
Using (\ref{basis}) and identifying
\begin{equation}
 g_7 = \int_{X_7} \widetilde G_7
\label{tildeg7}
\end{equation}
and
\begin{equation}
M_{ij} = - \frac14 \int_{X_7} \phi^i \wedge d \phi^j,
\label{Mdefint}
\end{equation}
gives (\ref{genW}).
By definition $M_{ij}$ is symmetric and $M_{ii}=0, \; \forall i$.
For our setup of orbifolds of twisted tori, the expansion of the
differential of the basis of 3-forms is
\begin{equation}
d \phi^i = -4 M_{ij} \widetilde \phi^j.
\label{dphii}
\end{equation}
Then,  consistency of exterior differentiation and the constancy of
$M_{ij}$ with respect to the internal coordinates implies
\begin{equation}
d \widetilde \phi^i  = 0.
\label{dtildephii}
\end{equation}
This in turn gives the closedness of the coassociative 4-form, in
agreement with (\ref{dstarphi2}).

\subsubsection{Supersymmetric Minkowski vacua}

Generic supersymmetric vacua are obtained for $D_i W = 0$.
This results in a negative semi-definite value of the cosmological constant
at the critical point
\begin{equation}
V_{*} = - 3 e^{K}|{\cal W}|^{2} \leq 0.
\label{Vcrit}
\end{equation}
It is clear that in order to obtain a supersymmetric Minkowski vacuum
the vanishing of (\ref{Vcrit}) must be imposed, which implies ${\cal
W} = 0$.
Altogether, the conditions for supersymmetric Minkowski vacua are
${\cal W} = 0 = \partial_{i}{\cal W}$.

The above conditions applied to (\ref{genW}) translate into the
following set of equations
\begin{equation}
\begin{array}{rcl}
M_{ij}t^{j} & = & 0,  \\[2mm]
M_{ij}\tau^{j} & = & -G_i,  \\[2mm]
\displaystyle \frac12 M_{ij}\tau^{i}\tau^{j} & = & -g_7.  
\end{array}
\label{solMink}
\end{equation}
The first one tells us that $t^i={\rm Re}\; T^{i}$ should be a null
eigenvector of $M$ and that $M$ must be therefore degenerate to have
solutions.
Also, there must be at least one null eigenvector, which has either
only positive components, so that $t_i > 0$, or four of them positive
and two negative.

Since $M$ has reduced rank, some of the equations in the second line
of (\ref{solMink}) are linearly dependent.
This implies that there would be some additional consistency
conditions between the fluxes $G_i$ and that some of the moduli
$\tau_i$ will be left unfixed.
Similarly, if a set of ${t_i}$ is a null vector of $M$, then also
${\lambda t_i}$ is a null vector and hence there is at least one
unfixed geometric modulus.
These facts are in accordance with the general expectation that
supersymmetric Minkowski vacua can never lead to complete moduli
stabilization without taking into account non-perturbative effects
\footnote{We would like to thank J.-P.~Derendinger for discussions on
this issue.}.

Using now the basis (\ref{basis}) and the subsequent formulas linking
the quantities in the superpotential to the geometric structure of the
internal manifold, we can also prove that supersymmetric Minkowski
vacua are obtained if and only if the internal manifold has
G$_2$-holonomy.
Using the expression of the G$_2$-form $\Phi = t_i \phi^i$ and of its
dual 4-form $\star \Phi = \frac{V}{t^i} \widetilde \phi^i$, we can
compute the torsion classes taking also into account (\ref{dphii}) and
(\ref{dtildephii}).
On the vacuum, where (\ref{solMink}) is satisfied, we find
\begin{eqnarray}
d \Phi &=& -4 \, t^i M_{ij} \widetilde \phi^j = 0,
\label{dphiMink}\\[2mm]
d \star \Phi &=&  d\left(\frac{V}{t^i} \widetilde \phi^i \right)= 0.
\end{eqnarray}
Here we have used the fact that the differential acts only on the
internal coordinates, so that $V$ and $t^i$ are constant.
The outcome is that Minkowski supersymmetric vacua require twisted
7-tori with G$_2$-holonomy.
The converse is also true as the $\phi^i$ span a basis of the 3-forms,
and therefore (\ref{dphiMink}) implies the first supersymmetry
condition in (\ref{solMink}).

\subsubsection{Supersymmetric AdS vacua}

These are determined by the vanishing of the K\"ahler covariant
derivatives of the potential (\ref{genW}):
\begin{equation}
D_i {\cal W} = M_{ij} T^j + i G_i -\frac{1}{T^i + \overline T^i} {\cal W} = 0.
\label{DiW}
\end{equation}
It is straightforward to see from (\ref{DiW}) that if ${\cal W}$ does
not depend on one or more moduli then  the only allowed supersymmetric vacua
are Minkowski as $D_i {\cal W} =\displaystyle -\frac{1}{T^i +
\overline T^i} {\cal W} = 0$.
The consequence is that we can find AdS vacua only if the potential
depends on all 7 moduli.

We can further strengthen this condition by separating the real and
imaginary pieces of (\ref{DiW}) as
\begin{eqnarray}
M_{ij}t^j & =& \frac{1}{2t^i} {\cal W}_{Re},  \label{reDW}\\
M_{ij}\tau^j + G_i& =& \frac{1}{2t^i} {\cal W}_{Im},
\label{imDW}
\end{eqnarray}
where we have introduced the real and imaginary parts of the
superpotential:
\begin{eqnarray}
{\cal W}_{Re} & = & \frac12 t^i M_{ij} t^j - \frac12 \tau^i M_{ij} 
\tau^j - G_i \tau^i + g_7  \label{Wr}\\
{\cal W}_{Im} & = & t^i\left(M_{ij}\tau^j + G_i\right). \label{Wi}
\end{eqnarray}
Now, summing over all the imaginary parts (\ref{imDW}) with
coefficients $t^i \neq 0$, we obtain
\begin{equation}
\sum t^{i} \hbox{Im} \left(D_i {\cal W}\right) = 0,\qquad
\Leftrightarrow \qquad {\cal W}_{Im} = \frac{7}{2}{\cal W}_{Im},
\label{incon}
\end{equation}
which is obviously consistent only for 
\begin{equation}
{\cal W}_{Im} = 0.
\label{WI0}
\end{equation}
This condition is equivalent to
\begin{equation}
M_{ij}\tau^j + G_i = 0,
\label{cond1}
\end{equation}
which also solves identically (\ref{imDW}) when ${\cal W}_{Im} = 0$.

The fact that supersymmetric AdS vacua can be obtained only for real
values of the superpotential at the critical point imposes further
constraints on the possible matrices $M$.
Indeed, if the quadratic part of ${\cal W}$ does not depend on some of
the moduli, i.e.~if $M$ is zero in some of the rows, then no
supersymmetric AdS critical points are allowed.
This can be shown by considering (\ref{DiW}) in these directions,
where one finds that
\begin{equation}
D_i {\cal W} = i G_i - \frac{1}{T^i + \overline T^i} {\cal W} = 0,
\label{D0W}
\end{equation}
implying that ${\cal W}$ should be purely imaginary.
Then we have ${\cal W}=0$ and due to (\ref{WI0}) we are back at the
Minkowski case.

The conclusion is that, in order to obtain supersymmetric AdS
solutions, the quadratic part of the superpotential should depend on
all 7 moduli.
In this case complete moduli stabilization is in principle possible,
provided that the additional constraints on $M$ coming from the Jacobi
identities (\ref{const}) are satisfied.
The vacua are identified by
\begin{equation}
\begin{array}{rcl}
M_{ij} t^j &=& \displaystyle \frac{1}{2 t^i} \left(\frac23 G_i 
\tau^i-\frac16 g_7\right), \\[2mm]
M_{ij} \tau^j + G_i &=& 0.
\end{array}
\label{solAdS}
\end{equation}

Once again we can use these conditions to check the form of the 
allowed G$_2$-structure.
The exterior differential on the G$_2$-form and its dual now give 
\begin{eqnarray}
d \Phi &=& -4\,t^i M_{ij} \widetilde \phi^j = \frac{\lambda}{t^i} 
\widetilde 
\phi^i \sim \star \Phi = \frac{V}{t^i} \widetilde \phi^i,
\label{dphiAdS}\\[2mm]
d \star \Phi &=&  d\left(\frac{V}{t^i} \widetilde \phi^i \right)= 0,
\end{eqnarray}
where we used (\ref{reDW}), the expansion of the dual 4-form $\star
\Phi$, and again the fact that we are taking the differential only
with respect to the internal coordinates, so that $V$ and $t^i$ are
constant.
We see that on the vacuum the only non-zero intrinsic torsion class is
$W_1 \sim \frac{\lambda}{V}$, i.e.
the manifold has weak G$_2$-holonomy.
The outcome is that supersymmetric AdS$_4$ vacua require twisted
7-tori with weak G$_2$-holonomy.
Finally, we can go backwards and show that having weak G$_2$-holonomy
implies the supersymmetry condition (\ref{reDW}), and hence a
supersymmetric AdS$_4$ solution.

\subsection{Examples}

In order to provide examples of vacua that satisfy the general
supersymmetry conditions derived in the previous section, we have
first to identify the possible matrices $M$ satisfying the Jacobi
constraints (\ref{const}) and then analyse the corresponding
superpotentials.
These constraints are required to give vacua of the effective
potential, which have a well defined interpretation in terms of
11-dimensional geometries.
Weaker constraints, such as consistency of the effective theory, may
lead to additional vacua, which however do not correspond to
compactifications of M-theory.
An example of this phenomenon is the AdS$_4$ vacuum of
\cite{Derendinger:2004jn}, that fixes all the moduli of the
4-dimensional theory.
Although this is a consistent truncation of ${\cal N} = 4 $ gauged
supergravity, it does not satisfy the 10-dimensional Jacobi identities
corresponding to (\ref{const}) 
\cite{Villadoro:2005cu}.

The conditions (\ref{const}) give very strong constraints on the
possible terms allowed in the superpotentials.
In order to derive consistent sets of $\tau_{IJ}^{K}$ we can use
the fact that (\ref{const}) are just 
the Jacobi identities for the
group-manifold on which, after taking the quotient with a discrete
subgroup, we  compactify M-theory.
This leads to the requirement that $\tau_{IJ}^{K}$ 
are the structure constants
of a 7-dimensional algebra whose adjoint is identified with the
fundamental of $sl(7)$ \cite{DallAgata:2005ff}.
All possibilities consistent with the orbifold projection can then be
analysed.
It turns out that they fall in four main categories:
\begin{itemize}

\item $SO(p,q) \times U(1)$, for $p+q =4$, 

\item $SO(p,q) \rtimes {\mathbb R}^4$, for $p+q =3$,

\item a 2-step nilpotent\footnote{A Lie algebra $g$ is called $n$-step
nilpotent when its lower central series $g^{(k+1)}=[g^{(k)},g], \;
g^{(0)}=g$ terminates at $g^{(n)}=0$ while it is called solvable when
its derived series $g_{(k+1)}=[g_{(k)},g_{(k)}], \; g_{(0)}=g$
terminates for some $k$.
Obviously nilpotency implies solvability.
} (metabelian) algebra $N_{7,3}$ ,

\item a solvable algebra $S_{6} \rtimes U(1)$ (which contains the
flat groups of \cite{Scherk:1979zr}).

\end{itemize}
We will see that the first and third lead to superpotentials that have
a quadratic part depending on all moduli, while the others do not.
This means that only these groups may lead to supersymmetric AdS vacua
and complete moduli stabilization, whereas the others can only lead to
supersymmetric Minkowski vacua.
The last two are allowed also when taking the orbifold
(\ref{orbifold}) to act freely.

\subsubsection{$SO(p,q) \times U(1)$ with $p+q = 4$}

The generic form of the matrix $M$ corresponding to this choice of
structure constants has a block form (possibly after an index
reshuffling):
\begin{equation}
M = \left(
\begin{array}{cc}
0_3 & A  \\[2mm]
{}^{t}A & 0_{4}   
\end{array}
\right),
\label{Mblock}
\end{equation}
where the matrix $A$ depends on 6 parameters and reads
\begin{equation}
A = \left(
\begin{array}{cccc}
\displaystyle  a_4 \frac{a_1}{a_2} &\displaystyle a_5 \frac{a_1}{a_3} 
& \displaystyle -a_6 \frac{a_1}{a_2} & a_1 
\\[4mm]
a_4 & \displaystyle - a_5 \frac{a_2}{a_3} & a_6 & a_2 \\[2mm]
\displaystyle -a_4 \frac{a_3}{a_2} & a_5 &\displaystyle a_6 \frac{a_3}{a_2} & a_3
\end{array}
\right).
\label{Aform}
\end{equation}
The corresponding structure constants can be used to 
build the algebra $so(p,q) \times
u(1)$
\begin{equation}
[H_a, H_b] = f_{abc} H_c, \quad [N_a, N_b] = g_{abc} N_c, \quad 
[G, H_a] = [G, N_a] = 0,
\label{sopq}
\end{equation}
where $f_{abc}$ and $g_{abc}$ can be the structure constants of
$SU(2)$ or $SL(2,{\mathbb R})$.
This depends on the actual values of the parameters $a_I$, since the
generators $H_a$, $N_a$ and $G$ are defined in terms of the $X_I$
generators satisfying $[X_I, X_J] = \tau_{IJ}^{K} X_K$ as
\begin{equation}
\begin{array}{c}
\displaystyle 
H_1 \equiv \pm \frac{1}{\sqrt{a_4 a_5}} X_5 \pm \frac{1}{\sqrt{a_3 a_6}} X_6, 
\quad H_2 \equiv \mp \frac{1}{\sqrt{a_1 a_5}} X_7 \mp \frac{a_2}{\sqrt{a_1 a_3 a_4 
a_6}}X_8, 
\\[3mm] 
\displaystyle 
H_3 \equiv \mp \sqrt{\frac{a_3}{a_1 a_5 a_6}}X_9 \pm \frac{1}{\sqrt{a_1 
a_4}}X_{10}, \\[4mm]
\displaystyle
N_1 \equiv \mp \frac{1}{\sqrt{a_4 a_5}} X_5 \pm \frac{1}{\sqrt{a_3 a_6}} X_6, 
\quad N_2 \equiv \pm \frac{1}{\sqrt{a_1 a_5}} X_7 \mp \frac{a_2}{\sqrt{a_1 a_3 a_4 
a_6}}X_8 ,
\\[3mm] 
\displaystyle
N_3 \equiv \pm  \sqrt{\frac{a_3}{a_1 a_5 a_6}}X_9  \pm \frac{1}{\sqrt{a_1
a_4}}X_{10}, \\[4mm]
G \equiv X_{11}.
\end{array}
\label{defgenerators}
\end{equation}

In the generic case with all 6 parameters $a_I \neq 0$, this matrix
gives a superpotential which has a quadratic part depending on all
moduli.
This is a necessary condition for obtaining supersymmetric AdS vacua
with complete moduli fixing.
Furthermore, it is also degenerate, which is the necessary condition
to obtain supersymmetric Minkowski vacua.
Unfortunately it can be checked that it does not allow for any
supersymmetric vacuum, either AdS or Minkowski.
The first type of vacua can only be excluded after an explicit
computation of (\ref{solAdS}).
These imply that there is no solution unless the superpotential
vanishes, which implies vanishing cosmological constant.
The flat vacua are, however, also excluded because all the null
eigenvectors have some vanishing components.
These are unacceptable, as they represent singular values of the real
parts of the moduli fields.

If some of the torsions are set to zero, we can obtain supersymmetric
Minkowski vacua with partial moduli stabilization.
An instance is given by the choice $a_{3} = a_4 = a_5 = 0$, which
represents a singular limit of the algebra defined by the generators
(\ref{defgenerators}), but is a perfectly well defined choice of the
matrix (\ref{Aform}) and thus of (\ref{Mblock}).
We will see later that this case is the intersection of this group
choice with the one leading to flat groups, which explains the result.

A different subgroup contained in this case is the direct product of
two copies of the Heisenberg algebra.
This, however, does not lead to supersymmetric vacua as well.

Finally, the choice $a_1 = 0$ and $a_3 = a_4 = a_5 = a_6 = 1$, $a_2 =
-1$ gives non-supersymmetric Minkowski vacua.
In this case the non-trivial commutators between the generators
are
\begin{equation}
\begin{array}{rclrcl}
[X_5, X_{10}] &=& X_7, & [X_5, X_7] &=& -X_{10}, \\[2mm]
[X_5, X_{9}] &=& X_8, & [X_5, X_8] &=& -X_{9}, \\[2mm]
[X_6, X_{8}] &=& X_{10}, & [X_6, X_{10}] &=& -X_{8}, \\[2mm]
[X_6, X_{7}] &=& X_9, & [X_6, X_9] &=& -X_{7}, \\[2mm]
\end{array}
\label{2flat}
\end{equation}
therefore forming a flat group \cite{Scherk:1979zr}.
The group described by (\ref{2flat}) constitutes actually  two smaller
copies of the one we analyse in \ref{secflatgroup}
and we refer the reader there for more details.

\subsubsection{$SO(p,q) \rtimes {\mathbb R}^4$ with $p+q = 3$}

This choice leads to a degenerate matrix that depends on six
parameters and (possibly after reshuffling of the indices) takes the
form:
\begin{equation}
M= \left( 
\begin{array}{ccccccc}
0 & b_1 \frac{b_2}{b_5} & b_1 \frac{b_3}{b_5} & b_1 \frac{b_4}{b_5}& 
b_1 & -2 b_1 \frac{b_4}{b_6} & 0\\
 b_1 \frac{b_2}{b_5} & 0&\frac{b_2 b_3 b_6}{b_4 b_5} & \frac{b_2 
 b_6}{b_5} & -2 \frac{b_2 b_6}{b_4} &\frac{b_2 b_3 b_6}{b_4 b_5} &0\\
 b_1 \frac{b_3}{b_5} &\frac{b_2 b_3 b_6}{b_4 b_5} &0 & -2 \frac{b_3 b_6}{b_5}& \frac{b_3 
 b_6}{b_4} & b_3 & 0\\
 b_1 \frac{b_4}{b_5} & \frac{b_2 
 b_6}{b_5} & -2 \frac{b_3 b_6}{b_5}& 0 &b_6& b_4 & 0\\
b_1 &-2 \frac{b_2 b_6}{b_4}&\frac{b_3 
 b_6}{b_4} & b_6 &0 & b_5 & 0\\
 -2 b_1 \frac{b_4}{b_6}  & b_2 &b_3 & b_4&  b_5 & 0 & 0\\
0 &0&0&0&0 & 0 & 0
\end{array}
\right).
\label{Mexample2}
\end{equation}
The underlying algebra can be understood by identifying the three
$SO(p,q)$ generators with $X_5$, $X_8$ and $X_9$, while the ${\mathbb
R}^4$ is generated by linear combinations of the remaining generators
with coefficients that depend on the $b_I$ parameters.

Since the matrix $M$ contains a line of zeros, the quadratic part of
the superpotential does not depend on one modulus. Consequently, it
cannot lead to supersymmetric AdS$_4$ vacua. Furthermore,
supersymmetric Minkowski vacua are excluded
as well since generic $b_I \neq 0$ lead to a single null eigenvector,
which points in the direction 7.
This means that $t_1= \ldots=t_6 = 0$, which is once again a singular
limit.

\subsubsection{The 2-step nilpotent algebra $N_{7,3}$}

This choice leads to a quadratic superpotential, which depends on all
the moduli.
The corresponding matrix $M$ is simply chosen to have non-zero values
on one specific line, for instance $M_{1i} \neq 0$, and $M_{ij} = 0$,
with $i,j= 2\ldots 7$.
This means that once more it depends on six parameters.
It is a 2-step nilpotent algebra because the generators can be grouped
in two sets $G_a$, $a = 1,\ldots,4$ and $H_\alpha$, $\alpha = 1,2,3$,
with commutator relations described as
\begin{equation}
[G, G] = H, \quad [G, H] = 0, \quad [H,H] = 0.
\label{gencomm}
\end{equation}
This clearly implies that the algebra is nilpotent as any 2 commutator
operations annihilate any generator.
For special choices of the parameters this algebra contains the 3- or
5-dimensional Heisenberg algebras.

Thanks to the special form of the matrix $M$, it is easy to prove its
degeneracy and also that no supersymmetric AdS or Minkowski vacua are
allowed.
The conditions for supersymmetric AdS lead to the vanishing of the
cosmological constant, while the null eigenvectors of the matrix $M$
necessarily contain once more vanishing components.
We also checked that no Minkowski vacua are allowed even when
considering complete supersymmetry breaking.

\subsubsection{Flat groups}
\label{secflatgroup}

The last possibility is given by matrices containing at least 3 lines
of zeros.
We will show that these include the flat groups first described in
\cite{Scherk:1979zr}.
These matrices can be singled out by choosing the non-vanishing
geometric fluxes to have the form $\tau_{X I}^{J}$, where $X$ stands
for one of the indices $5,\ldots,11$ and $I,J$ range over the
remaining ones (the flat groups are the special subcase when these
are in addition antisymmetric in $IJ$).

Given the orbifold projections ({\ref{orbifold}), we select a $4\times
4$ matrix $M$, which, by an appropriate choice of  elements, can be
made degenerate.
For instance, choosing $X=5$, the only non-trivial twists are
$\tau_{56}^{11}=-k_1$, $\tau_{510}^7=k_2$, $\tau_{59}^8=k_3$,
$\tau_{58}^9=-k_4$, $\tau_{57}^{10}=-k_5$, $\tau_{511}^6 = k_6$.
One can easily check that the Jacobi identities are true for these
sets of twists for any value of the $k$'s.
The quadratic part of the superpotential then depends on 4 moduli
$T_2, T_3, T_4, T_5$ and the corresponding matrix reads
\begin{equation}
M = \left(\begin{array}{cccc}
0 & k_1 & k_2 & k_3  \\
k_1 & 0 & k_4 & k_5  \\
k_2 & k_4 & 0 & k_6  \\
k_3 & k_5 & k_6 & 0   
\end{array}
\right).
\label{Mflat}
\end{equation}

Since the superpotential (\ref{Mflat}) depends on only 4 moduli and
the related potential is therefore of the no-scale type
\cite{Cremmer:1983bf}, the only vacua to be expected are flat
Minkowski space--times.
Using the first condition in (\ref{solMink}), it is clear that such
vacua can be obtained when
\begin{equation}
t_1 k_1 = t_6 k_6, \quad t_4 k_4 = t_3 k_3, \quad t_5 k_5 = t_2 k_2, 
\label{vacminkflat}
\end{equation}
and $t_1 k_1 \pm t_2 k_2 \pm t_3 k_3 = 0$.
The different signs lead to the same manifold up to reshuffling of the
vielbeins.
It is also clear that a rescaling of the twisting parameters can be
reabsorbed in a rescaling of the size moduli. Therefore
these different choices do not lead to different internal manifolds
either.
For this reason, in the following we focus on the case given by
the flat groups of \cite{Scherk:1979zr}.
These are obtained for $k_4= k_3$, $k_6 = k_1$ and $k_5 = k_2$ and in
addition we impose $k_1+k_2+k_3=0$, so that the matrix $M$ becomes
degenerate.
This choice of parameters implies that (\ref{Mflat}) has a null
eigenvector given by $\lambda \{1,1,1,1\}$.

The matrix degeneracy has a special meaning in terms of the geometry
of the corresponding internal manifold.
Computing the torsions (\ref{dphi}) and (\ref{dstarphi}) for the flat
group at $t_2=t_3=t_4=t_5=1$, we obtain\footnote{We use the shorthand
notation $\eta^{I_1 \cdots
I_n}=\eta^{I_1}\wedge\cdots\wedge\eta^{I_n}$.}
\begin{equation}
d\Phi = \left(k_1+k_2+k_3\right)\left(\eta^{58 10 11}-
\eta^{57911}+\eta^{5678}+\eta^{56910}\right), \quad  d\star\!\Phi = 0,
\label{dphiflat}
\end{equation}
which vanishes for $k_1+k_2+k_3=0$, i.e.~the internal manifold has
G$_2$-holonomy.
This is not a complete surprise, since the corresponding twisted tori
have vanishing Riemann tensor and therefore they are flat (hence the
name flat groups).

{}For this reason, we expect that the potential obtained using $M$
admits a supersymmetric Minkowski vacuum with all  fluxes set to
zero.
Indeed, the choice $k_1 = -k_2-k_3$ fulfills the first requirement to
get to a supersymmetric Minkowski vacuum: a degenerate matrix $M$.
Furthermore, the fluxes should satisfy $G_1=G_6=G_7=G_2+G_3+G_4+G_5=0$
as a consequence of the second equation in (\ref{solMink}).
As expected there is a solution with all fluxes set to
zero and with the moduli taking values  $t_2 = t_3 = t_4 = t_5 =
\lambda>0$ and $\tau_2 = \tau_3 = \tau_4 = \tau_5 = 0$ while $T_1,T_6,
T_7$ are left unfixed.
This is the standard Minkowski background found upon compactification
on a G$_2$-holonomy manifold.

The metric of the space resulting from quotients of the flat groups
${\cal G}$ can be derived by constructing the right-invariant
vielbeins $\eta^I$, that satisfy (\ref{torsion}) with the flat-group 
structure constants.
These are obtained from the right-invariant Maurer--Cartan forms
$\Omega = dg g^{-1}$, where $g \in {\cal G}$ is a group 
representative (see \cite{Hull:2005hk}).
The vielbeins obtained from $\Omega$ are 
\begin{equation}
\begin{array}{rclcrclcrcl}
\eta^5 &=& dy^5, &\quad & \eta^6 &=& dy^{6} - k_1 y^{11} dy^5, &\quad&  \eta^7 
&=& dy^{7} - k_2 y^{10} dy^5,\\[2mm]
\eta^8 &=& dy^8 -k_3 y^9 dy^5, &\quad & \eta^9 &=& dy^{9} + k_3 y^{8} dy^5, 
&\quad&  \eta^{10} 
&=& dy^{10} + k_2 y^{7} dy^5,\\[2mm]
 && &\quad & \eta^{11} &=& dy^{11} + k_1 y^{6} dy^5, &\quad& &&
\end{array}
\label{ds2flat}
\end{equation}
and give the twisted torus metric $ds^2 = \sum_{I} \eta^{I} \otimes 
\eta^{I}$.
This space is made compact by imposing the following identifications
\begin{equation}
\begin{array}{rcl}
y^{5}\sim y^5 +1, & &\\[2mm]
\left\{
\begin{array}{rcl}
y^6 &\sim& y^6 + \cos \left(k_1 y^5\right)\\
y^{11} &\sim& y^{11} - \sin \left(k_1 y^5\right)
\end{array}
\right. &&
\left\{\begin{array}{rcl}
y^6 &\sim& y^6 + \sin \left(k_1 y^5\right)\\
y^{11} &\sim& y^{11} + \cos \left(k_1 y^5\right)
\end{array}
\right. \\[4mm]
\left\{
\begin{array}{rcl}
y^7 &\sim& y^7 + \cos \left(k_2 y^5\right)\\
y^{10} &\sim& y^{10} - \sin \left(k_2 y^5\right)
\end{array}
\right. &&
\left\{\begin{array}{rcl}
y^7 &\sim& y^7 + \sin \left(k_2 y^5\right)\\
y^{10} &\sim& y^{10} + \cos \left(k_2 y^5\right)
\end{array}
\right. \\[4mm]
\left\{
\begin{array}{rcl}
y^8 &\sim& y^8 + \cos \left(k_3 y^5\right)\\
y^{9} &\sim& y^{9} - \sin \left(k_3 y^5\right)
\end{array}
\right. &&
\left\{\begin{array}{rcl}
y^8 &\sim& y^8 + \sin \left(k_3 y^5\right)\\
y^{9} &\sim& y^{9} + \cos \left(k_3 y^5\right),
\end{array}
\right. 
\end{array}
\label{identif}
\end{equation}
under which the vielbeins and hence the metric are globally well-defined.

These identifications follow from the right quotient of ${\cal G}$ 
with its discrete subgroup $\Gamma = {\cal G}({\mathbb 
Z})$. One can also construct the right-invariant vector fields
\begin{equation}
\begin{array}{rcl}
X_5 &=& \partial_5 + k_1 \left(y^{11} \partial_6 - y^6 
\partial_{11}\right) + k_2 \left(y^{10} \partial_7 - y^7 
\partial_{10}\right)+ k_3 \left(y^9 \partial_8 - y^8 
\partial_{9}\right), \\[2mm]
 X_I &=& \partial_I, \qquad I = 6,\ldots,11.
\end{array}
\label{Killright}
\end{equation}
which generate the flat group algebra
\begin{equation}
[ X_{I}, X_J] = -\tau_{IJ}^K  X_K.
\label{leftalgebra}
\end{equation}
Since these vector fields are right-invariant, they are 
well-defined on the 
discrete quotient defined by the right action of $\Gamma$ on 
${\cal G}$ and generate the left action on ${\cal G}/\Gamma$.
However, not all of them are actual isometries of the
compact space ${\cal G}/\Gamma$.
Only the generators that lie in the commutant of ${\cal G}\equiv{\cal G}({\mathbb
R})$ in ${\cal G}({\mathbb Z})$ are isometries of the compact space.
This implies that the generators $ X_I$, for $I=6,\ldots,11$
are not Killing vectors of the compact metric. Hence only $ X_5$
generates an isometry of the metric  $ds^2 = \sum_{I} \eta^{I} \otimes 
\eta^{I}$.

The twisted torus we are describing is a fibration along $y^5$ of
three 2-tori, parametrized by the couples of coordinates $\{y^6,
y^{11}\}$, $\{y^7, y^{10}\}$, $\{y^8, y^{9}\}$.
Moreover, the identifications in (\ref{identif}) show that this 
fibration is nothing but a rotation of these tori around $y^5$ by 
angles depending on $k_1$, $k_2$ and $k_3$.
Because of this twist, translations in the coordinates of the three 
2-tori are not isometries while the vector field $ X_5$, 
that is the sum of the translation along $y^5$ with the angular 
shifts on the tori, is a Killing vector.

The Riemann tensor 
of the metric  $ds^2 = \sum_{I} \eta^{I} \otimes 
\eta^{I}$ is vanishing for any value of the twist parameters
$k_i$.
This means that ${\cal G}/\Gamma$  is nothing but flat space, and, in the
compact version, flat ${\mathbb T}^{7}$.
This explains why the resulting effective theory yields consistent
Minkowski vacua (supersymmetric and  non supersymmetric).
There is actually an alternative (and equivalent) description of
${\cal G}/\Gamma$ which reproduces explicitly the flat space metric.
This can be obtained by implementing the change of coordinates
\begin{equation}
\begin{array}{rclrcl}
y^6 &=& \sin (k_1 x^5 ) x^{6} + \cos( k_1 x^5)
x^{11}, &  y^{11} &=&\cos( k_1 x^5 ) x^{6} -  \sin (k_1 x^5)
x^{11}, \\[2mm]
y^7 &=&\sin (k_2 x^5 ) x^{7} + \cos( k_2 x^5)
x^{10},&   y^{10} &=& \cos( k_2 x^5) x^{7} - \sin (k_2 x^5 )
x^{10}, \\[2mm]
y^8 &=& \sin (k_3 x^5 ) x^{8} + \cos( k_3 x^5)
x^{9}, &  y^{9} &=& \cos(k_3 x^5) x^{8} - \sin (k_3 x^5 )
x^{9}, \\[2mm]
&& & y^5 &=& x^5. 
\end{array}
\label{changeofcoords}
\end{equation}
The resulting Euclidean metric $ds^{2} = \sum_I dx^I \otimes
dx^I$ has the standard identifications 
$x^I\sim x^{I} + 1$ for $\;I=6,\ldots,11$.  For $x^5$, however, the identification
reads
\begin{equation}
\begin{array}{rcl}
x^{5} &\sim& x^{5} + 1, \\[2mm]
x^6 &\sim& \cos (k_1) x^6 - \sin(k_1 ) x^{11}, \\[2mm]
x^7 &\sim& \cos (k_2) x^7 - \sin(k_2 ) x^{10}, \\[2mm]
x^8 &\sim& \cos (k_3) x^8 - \sin(k_3) x^{9}, \\[2mm]
\end{array}
\quad
\begin{array}{rcl}
x^9 &\sim& \sin (k_3 ) x^8 + \cos(k_3 ) x^{9}, \\[2mm]
x^{10} &\sim& \sin (k_2 ) x^7 + \cos(k_2 ) x^{10}, \\[2mm]
x^{11} &\sim& \sin (k_1 ) x^6 + \cos(k_1) x^{11}.
\end{array}
\label{identif2}
\end{equation}
This reduces to the ordinary one when the twisting coefficients $k_i$ are 
chosen to be integers\footnote{We have defined the sine and cosine 
functions with period 1.}.

This description can also be obtained by introducing the 
left-invariant vielbeins $\widetilde \eta^{I}$, constructed from the
left-invariant Maurer--Cartan form $\widetilde \Omega = g^{-1}dg$, and
satisfying
\begin{equation}
d \widetilde \eta^{I} + \frac12 \tau_{JK}^{I} \widetilde \eta^{J} \wedge 
\widetilde \eta^{K}=0.
\label{leftinvviel}
\end{equation}
It should be noted that the sign in (\ref{leftinvviel}) is the 
opposite of (\ref{torsion}).
These vielbeins are given  by
\begin{equation}
\begin{array}{rclcrcl}
\widetilde \eta^5 &=& dx^5, &\quad & \widetilde \eta^6 &=&  \cos (k_1 
x^5) 
dx^{6} -\sin(k_1 x^5) dx^{11},\\[2mm]
\widetilde \eta^7 
&=& \cos( k_2 x^5) dx^{7} -\sin ( k_2 x^5) dx^{10}, &\quad&
\widetilde \eta^8 &=& \cos(k_3 x^5) dx^8 -\sin(k_3 x^5) dx^9,\\[2mm]
\widetilde \eta^9 &=& \sin(k_3 x^5) dx^{8} +\cos(k_3 x^5) dx^{9}, 
&\quad&  \widetilde \eta^{10} 
&=& \sin( k_2 x^5)dx^{7} +\cos( k_2 x^5) d x^{10},\\[2mm]
 \widetilde \eta^{11} &=& \sin(k_1 x^5) dx^{6} +\cos(k_1 x^5) dx^{11},   &\quad &&&
\end{array}
\label{leftviel}
\end{equation}
and the metric $ds^2 = \sum_{I} \widetilde\eta^{I} \otimes 
\widetilde\eta^{I}$ is precisely 
the flat Euclidean metric, as it can be clearly seen by the fact that the left-invariant
vielbeins
are $x^5$-rotations of the ordinary straight differentials.

The global identifications in this case correspond to the left quotient 
$\Gamma
\backslash {\cal G}$ of ${\cal G}$ 
by $\Gamma$. The right-action is generated by the left-invariant vectors fields
\begin{equation}
\begin{array}{rclrcl}
\widetilde X_6 &=& \cos(k_1 x^5) \partial_{6} - \sin(k_1 x^5 )
\partial_{11}, &  \widetilde X_{11} &=& \sin (k_1 x^5) \partial_{6}
+ \cos (k_1 x^5) \partial_{11}, \\[2mm] 
\widetilde X_7 &=& \cos(k_2 x^5)
\partial_{7} - \sin(k_2 x^5) \partial_{10},&  \widetilde X_{10} &=& \sin
(k_2x^5) \partial_{7} + \cos(k_2 x^5) \partial_{10}, \\[2mm] 
\widetilde X_8 &=&
\cos(k_3 x^5 )\partial_{8} - \sin(k_3 x^5) \partial_{9}, &  \widetilde X_{9}
&=& \sin (k_3x^5) \partial_{8} + \cos(k_3 x^5) \partial_{9}, \\[2mm] 
&& & \widetilde X_5 &=& \partial_{5}, 
\end{array}
\label{leftKilling}
\end{equation}
and it can be checked that they indeed satisfy
the flat-group algebra 
\begin{equation}
[\widetilde X_I , \widetilde X_J] = \tau_{IJ}^{K} \widetilde X_K .
\label{rightalg}
\end{equation}
Also in this case we can see explicitly that only $\widetilde X_5$ is generating an
isometry of $\Gamma
\backslash {\cal G}$.

Given this change of coordinates we can now understand why the
Scherk--Schwarz reduction on this twisted torus (that is just an
ordinary flat torus for integer $k_i$) fixes some moduli.
The ordinary truncation of the Kaluza--Klein spectrum expansion on the
flat ${\mathbb T}^7$ is done by keeping only the zero modes of the
Laplace--Beltrami operators, i.e.~the moduli $t_I$ are associated to
harmonic 3--forms.
Instead, the Scherk--Schwarz truncation selects a different set of
moduli.
These are the singlets under the action of the symmetry group
generated by either $X_L$ or $\widetilde X_R$ according to the choice of quotient
${\cal G}/\Gamma$ or $\Gamma \backslash {\cal G}$ respectively.
In terms of the Kaluza--Klein modes this implies that only some (or
none) of the original zero-modes are kept, while massive modes are
introduced in the truncated spectrum in a way that is consistent with
the higher-dimensional equations of motion.
This has to be opposed to the usual Kaluza--Klein zero-mode truncation
which in general is not consistent on group manifolds or cosets
\cite{Hull:2005hk}.

In order to illustrate  further the above points, let us compare
the ans\"atze for the reduction of the metric in the two cases.
The ordinary truncation of the Kaluza--Klein expansion on a generic
twisted torus would be
\begin{equation}
ds^{2} = g_{\mu\nu}(x) dx^{\mu} \otimes dx^{\nu} + \left(G_{IJ}(y) +
g_{IJ}(x) \right) \left(dy^{I} + K^{I}_{L}A_{\mu}^{L}dx^\mu\right) \otimes
\left(dy^{J} + K^{J}_{M}A_{\nu}^{M}dx^{\nu}\right),
\label{metrKK}
\end{equation}
where $G_{IJ}(y) dy^I \otimes dy^J$ is the metric of the twisted torus
and $K^{I}_{J}(y)$ are the Killing vectors of the internal isometries.
The Scherk--Schwarz truncation instead is essentially an expansion
around the (left-) right-invariant vielbeins
\begin{equation}
ds^{2} = g_{\mu\nu}(x) dx^{\mu} \otimes dx^{\nu} +
(\delta_{IJ}+g'_{IJ}(x)) \left(\eta^{I} + A_{\mu}^{I}dx^{\mu}\right) \otimes
\left(\eta^{J} + A_{\nu}^{J}dx^{\nu}\right).
\label{metrSS}
\end{equation}
It is clear that the usual Kaluza--Klein moduli $g_{IJ}(x)$ 
parametrize  
a different set of fluctuations  from the Scherk--Schwarz moduli
$g'_{IJ}(x)$. Rewriting (\ref{metrSS}) in a form comparable to
(\ref{metrKK}) and since $\eta^I=U^I_J(y) dy^J$, where $U^I_J(y)$ is the
twisting matrix,
reveals that the Scherk-Schwarz ansatz is not a truncation
to zero-modes but it rather corresponds to $y$-dependent fluctuations 
$g_{KL}(x,y)=g_{IJ}'(x)U^I_K(y) U^J_L(y)$.

For the flat group, where the twisted torus is actually flat and hence
$G_{IJ}=\delta_{IJ}$, the symmetry group $U(1)^{7}$ of the first
ansatz is realised as isometries of the vacuum, whereas for the
Scherk--Schwarz reduction obtained as the quotient of the non-compact
group ${\cal G}$ one cannot find a ${\cal G}$-invariant ground state 
and only a U(1) can be preserved. This U(1) is generated by the
unique Killing vector  we found earlier.

Coming back to the description of the effective light degrees of
freedom, we can see the mass generation in the following way.
In the ordinary Kaluza--Klein reduction on ${\mathbb T}^{7}$, 
one keeps the moduli associated to the
harmonic 3-forms so that the G$_2$-form is expanded as
\begin{equation}
\begin{array}{rcl}
\Phi_{\rm KK} &=& -t_1 dx^5 \wedge dx^6 \wedge dx^{11} + t_2 dx^6 \wedge dx^9 
\wedge dx^{10}+ t_3 dx^{6}\wedge dx^7 \wedge dx^8 \\[2mm]
&+& t_4 dx^8 \wedge 
dx^{10}\wedge dx^{11} 
- t_5 dx^{7} \wedge dx^9 \wedge dx^{11} - t_6 dx^5 \wedge dx^7 
\wedge dx^{10}\\[2mm]
&-& t_7 dx^5 \wedge dx^8 \wedge dx^9.
\end{array}
\label{Phiflat}
\end{equation}
It is clear that although the metrics for the ordinary ${\mathbb T}^7$
and for the left-quotient of the flat group look formally the same, the
forms that can be constructed using the straight differentials are not
invariant under the action of (\ref{identif2}) when
the $k_i$ are not integers. More generically, 
this is also true for the 3-forms appearing in the definition of the
G$_2$-structure. Therefore we cannot use the same expansion for
the Scherk--Schwarz reduction as for the truncation of the
Kaluza--Klein spectrum to the zero-modes.

All the above boil down to the fact that in the Scherk--Schwarz
reduction one has to define all invariant forms in terms of either the
left-invariant vielbeins $\tilde \eta^I$ or the right-invariant ones
$\eta^{I}$, and then promote the corresponding coefficients to
spacetime dependent moduli fields.
For example, for the left-quotient $\Gamma \backslash {\cal G}$ the
expansion of $\Phi$ would be
\begin{equation}
\begin{array}{rcl}
\Phi_{\rm SS} &=& -t_1 \widetilde \eta^5 \wedge \widetilde \eta^6 \wedge \widetilde \eta^{11} + t_2 \widetilde \eta^6 \wedge \widetilde \eta^9 
\wedge \widetilde \eta^{10}+ t_3 \widetilde \eta^{6}\wedge \widetilde \eta^7 \wedge \widetilde \eta^8 
+ t_4 \widetilde \eta^8 \wedge 
\widetilde \eta^{10}\wedge \widetilde \eta^{11}\\[2mm]
&-& t_5 \widetilde \eta^{7} \wedge \widetilde \eta^9 \wedge \widetilde \eta^{11} - t_6 \widetilde \eta^5 \wedge \widetilde \eta^7 
\wedge \widetilde \eta^{10}- t_7 \widetilde \eta^5 \wedge \widetilde 
\eta^8 \wedge \widetilde \eta^9.
\end{array}
\label{PhiSS}
\end{equation}
Only the three invariant 3-forms associated to $t_1$, $t_6$ and $t_7$
moduli correspond to the ones in (\ref{Phiflat}), i.e.~$\widetilde
\eta^5 \wedge \widetilde \eta^6 \wedge \widetilde \eta^{11} = dx^5
\wedge dx^6 \wedge dx^{11}$ etc.
Also, a fourth harmonic form can be obtained when $t_2 = t_3 = t_4 =
t_5$ and $k_1 + k_2 + k_3 = 0$.
Therefore, the corresponding fields remain massless moduli also in the
Scherk--Schwarz reduction, exactly as we found in the analysis of the
conditions leading to supersymmetric vacua.
The remaining 3-forms contain explicit dependence on the $x^5$
coordinate and are associated to non-harmonic 3-forms leading to
massive moduli.
In general, when the twisting coefficients $k_i$ are arbitrary, the
invariant G$_2$ form is not closed, because the rotations of the
vielbeins (\ref{leftviel}) with respect to the usual straight
differentials yields $x^5$-dependent factors of the form $\cos [(k_1 +
k_2 + k_3) x^{5}]$ or $\sin [(k_1 + k_2 + k_3) x^{5}]$ multiplying the
usual harmonic forms.

As far as the metric decomposition is concerned, one can easily see that
while ordinary toroidal 
Kaluza--Klein compactifications associate moduli to
coordinate rescalings and, therefore, to the differentials $R_I dx^I$,
the Scherk--Schwarz reductions associate moduli to the vielbeins
$R_I \eta^{I}$. 
The overall effect is similar to that responsible for the generation
of masses in the 1-dimensional Scherk--Schwarz reduction of a field
$\varphi(x,y)$ which possesses a global U(1) invariance $\varphi \to
{\rm e}^{i \alpha} \varphi$ \cite{Scherk:1978ta}.
In that case, the Scherk--Schwarz expansion
\begin{equation}
\varphi (x,y) ={\rm e}^{i \alpha y} \sum_{n} {\rm e}^{i n y} \varphi_{n}(x)
\label{SScircle}
\end{equation}
differs from the ordinary Kaluza--Klein by an additional factor
depending on the parameter $\alpha$.
Since an integer $\alpha$ corresponds to a relabeling of the modes
$\varphi_{n}(x)$, we should restrict $\alpha \in [0,1)$.
This way the effective mass is always  smaller than the
Kaluza--Klein masses.
In our flat group example the twisting coefficients $k_i$ are the
analogues of the parameter $\alpha$.
In the special instance that these coefficients are integers, the
identifications become those of the ordinary ${\mathbb T}^{7}$ and one
could use the ordinary Kaluza--Klein reduction, thereby getting seven
massless moduli.
On the other hand, the Scherk--Schwarz truncation keeps only modes
which are singlets under the symmetry group generated by either $X_L$
or $\widetilde X_R$ , thereby excluding all massless fluctuations of
the internal metric that are not invariant.

The upshot is that the Scherk--Schwarz truncation
correctly captures the low-energy degrees of freedom only when the
twisting coefficients are non-integers.
Presumably this is also true for more general twisted tori and not
just for the flat groups we considered here.
However, although in ordinary supergravity the deformation parameters
corresponding to the coefficients $\tau_{IJ}^K$ can be arbitrarily
small, string theory usually forces some quantization conditions
\cite{Hull:2005hk}.
Then, the efficiency of this moduli stabilization scheme would be
questionable.

{}From this discussion, it becomes once more apparent why these examples
allow for non-supersymmetric vacua.
Given that the manifold described by (\ref{ds2flat}) is just flat
space, it is a valid internal manifold for Minkowski vacua for any
value of the twisting parameters $k_i$.
Indeed, even in the case where the values of $k_i$ do not allow for a
supersymmetric Minkowski vacuum, they still allow a non-supersymmetric
one.
The potential is of the no-scale type and the minima are the Minkowski
vacua obtained at the same values of the moduli as those of the
supersymmetric critical points of the cases analysed above.
The interesting fact is that, despite being flat space, this choice of
twisting parameters implies that the holonomy group is not in G$_2$,
because $d\Phi \neq 0$.
This should have been expected as it is well-known that G$_2$-holonomy
implies supersymmetry.
The actual holonomy group ${\mathbb Z}_2\times {\mathbb
Z}_2\times{\mathbb Z}_2$ is now embedded in SO(7), but not in G$_{2}$,
because of the twists.

The last part of the discussion on the possible vacua for the flat
group manifolds is about the possibility of obtaining vacua with
4-form flux turned on.
We can actually show that more supersymmetric Minkowski vacua could be
found by turning on the 4-form flux in the directions containing the
index $5$, which means turning on precisely the linear terms in
(\ref{Wexpanded}) that depend on the 4 moduli $T_2, T_3, T_4, T_5$.
When this happens the potential gets new terms from the $4$-form
fluxes, which may fix the axions to a non-zero value, while the volume
moduli remain fixed at the same values as before.
This happens, for instance, when generating a potential from fluxes of
the form $-i \left(k_2 T_2 +k_3 T_3 + k_1 T_5\right)$.

Since the corresponding internal manifold still has G$_2$-holonomy,
this vacuum should be inconsistent as an 11-dimensional configuration
because of the no-go theorem of \cite{Maldacena:2000mw}.
What really happens is that this solution can still be considered a
consistent background, though trivially related to the previous one.
This is due to the fact that the fluxes that are turned on in this way
correspond to trivial fluxes on 4-forms, which are not dual to the
4-cycles when the twisting is performed.
As noted above, the introduction of the $\tau_{IJ}^{K}$ parameters and
the choice of performing a Scherk--Schwarz reduction imply that we
expand around non-harmonic 3-forms.
Actually, the 4-form flux leading to supersymmetric vacua is
cohomologically zero.
For example, the flux corresponding to the superpotential $-i
\left(k_2 T_2 +k_3 T_3 + k_1 T_5\right)$ is $g=-4 (k_2
\widetilde\phi^2+ k_3 \widetilde\phi^3+k_1 \widetilde\phi^5)=4
d\phi^4$, as we can see by using $\widetilde\phi^i=\frac{t_i^2}{V}
(\star\phi^i)$.

In the previous reduction we distinguished the non-trivial part $g$
and the trivial one, which is expressed as $\tau C$ and simply shifts
the values of the axions.
If, on the other hand, we let $g$ also have a trivial part, then the
consistent backgrounds are generically given by $G = g+\tau C$, which
is harmonic.
In our example this is the case because $G$ is identically vanishing.
The punchline is that this vacuum is indistinguishable from the
previous one, from the 11-dimensional point of view.
Also in the effective theory, the appearance of this type of flux in
the gauge algebra and in the Lagrangian can be removed entirely by
field redefinitions \cite{Hull:2005hk}.

If one introduces (cohomologically) non-trivial fluxes instead, then
one does not find any vacuum but rather a run-away potential, in
accordance with the no-go theorem \cite{deWit:1987xg,Maldacena:2000mw}.
This shows that the above vacuum, as well as many of those presented
in the literature, for instance those of \cite{DAuria:2005dd}, are a
trivial redefinition of the no-flux vacua.

The last comment concerns the possible corrections to these
backgrounds due to higher order derivative terms in the action.
Looking at $R^4$ terms, there are four types of possible corrections
(see for instance \cite{Vafa:1995fj,Lu:2003ze}).
Since the Riemann tensor is exactly vanishing on the vacuum, we can
see that these vacua are stable against these corrections.
When varying the contribution of the 
$R^{4}$ terms to the potential with respect
to the moduli, one still gets a critical point as its variation
vanishes for any value of the moduli that was already  unfixed and
without changing the
value of the ones that were fixed before.
It would be of course very interesting to reproduce explicitly these 
contributions as a
modification to the ${\cal N} = 1$ superpotential.

\section{Type IIA superpotentials}

We consider now the circle reduction of the potential (\ref{Wform}) to
type IIA theory.
In the first subsection we assume that (\ref{Wform}) has general
validity and we perform the reduction for a generic internal manifold
$X_7$ with an isometry.
In the second subsection we  reduce the twisted
7-torus when the choice of structure constants allows isometries.

\subsection{General reduction}

{}For a general fibered G$_2$-structure, the manifold $X_7$ is defined
as a circle fibration of an SU(3)-structure manifold $M_6$, with a
metric of the form \cite{Chiossi:2002kf}
\begin{equation}
ds_{X}^{2} = \alpha \otimes \alpha + \pi^{*} ds^2_{M},
\label{metricX7}
\end{equation}
where $\pi$ defines the fibration.
The reduction is performed by expressing the G$_2$-structure in terms
of the 6-dimensional SU(3)-structure as
\begin{eqnarray}
\Phi & = & (\pi^{*}J) \wedge \alpha + (\pi^{*}\rho), \label{Phifib}\\
\star \Phi & = & \frac12 (\pi^{*}J) \wedge (\pi^{*}J) + (\pi^{*}\hat \rho) 
\wedge \alpha. 
\label{starphifib}
\end{eqnarray}
Here $\rho$ and $\hat \rho$ are the real and imaginary parts of the
$(3,0)$-form $\Omega$.

In our case the 1-form $\alpha$ reads
\begin{equation}
\alpha = e^{\frac23 \phi}\left(dy^{11} + A + \sigma\right),
\label{alpha}
\end{equation}
with $d\sigma = f_{(2)}$, the background Ramond--Ramond (RR) 2-form
flux of type IIA theory.
Notice that $\sigma$ should not be globally defined on $M_6$, since we
would like a flux that is not pure gauge.
In addition, $\pi$ is defined so that the metric of the 6-dimensional
base is in the string frame.
In other words:
\begin{equation}
ds_{7}^2 = e^{-\frac{2\phi}{3}} ds^2_{6} + e^{\frac{4\phi}{3}}
(dy^{11} + C_1)^2,
\label{reduction}
\end{equation}
and we conclude that $\pi^{*} J = e^{-\frac23 \phi} J$ and $\pi^{*}
\rho = e^{-\phi} \rho$.
Finally, the 11-dimensional fields and background fluxes are reduced
accordingly:
\begin{equation}
C = {\cal C} + e^{-\frac23 \phi}B \wedge \alpha, \quad g = f_{(4)} + 
e^{-\frac23 \phi} h \wedge \alpha.
\label{redflux}
\end{equation}

Plugging these expressions in (\ref{Wform}) yields
\begin{equation}
4 {\cal W}_{IIA} = \frac12 \int J_c \wedge J_c \wedge (dA +f_{(2)}) + \int f_{(4)} \wedge J_c -
\int \left( dJ_c +\, h\right)\wedge \Omega_c,
\label{IIApot}
\end{equation}
where we have defined
\begin{equation}
\Omega_c =  {\cal C}+ i e^{-\phi}\rho, \quad J_c = B+ i J.
\label{omcJc}
\end{equation}
The 10-dimensional dilaton $\phi$ is related to the 4-dimensional one
$\phi_4$ as
\begin{equation}
\phi=\phi_4-\frac14 \log \det g_{(6)},
\label{dilrel}
\end{equation}
where $g_{(6)}$ is the metric of $M_6$.

The expression (\ref{IIApot}) of the type IIA superpotential agrees
with those presented in \cite{Grimm:2004ua,Camara:2005dc,
Villadoro:2005cu,House:2005yc,Grana:2005ny} upon appropriate
redefinitions of the SU(3)-structure.
Furthermore, in type IIA the superpotential can be further completed
with terms cubic in the moduli by considering the massive IIA theory.
Since there is no known lift of massive type IIA supergravity to
11-dimensional supergravity we cannot expect similar terms to arise
from the reduction of (\ref{Wform}).
Notice that in \cite{Hull:1998vy} massive type IIA {\em superstrings}
were obtained from M-theory by a Scherk--Schwarz reduction on a
${\mathbb T}^3$ that shrinks to zero size.
Unfortunately, this mechanism lies outside the region of validity of
our effective supergravity approach.

\subsection{Reduction of the twisted 7-torus}

The superpotentials we derived from twisted tori compactifications of
M-theory have the same form as the type IIA superpotentials obtained
from twisted tori compactifications.
They differ, however, in the number of independent terms.
M-theory yields a 4-dimensional ${\cal W}$ as in (\ref{Wexpanded}),
which contains more quadratic couplings between the moduli than those found
in the type IIA superpotentials.
This implies that either these couplings have to vanish when
considering reductions to type IIA supergravity from M-theory or that
the resulting type IIA backgrounds are not twisted tori.
These possibilities depend on the choices of isometries upon which one
reduces the 11-dimensional theory.

We have seen in the previous section that only some of the symmetries
of the Scherk--Schwarz ansatz for the reduction on twisted tori are
actual isometries of the possible internal metrics.
If, for instance, we look at the 2-step nilpotent algebra $N_{7,3}$,
we can obtain non-trivial 2-form fluxes, with a constant dilaton, upon
reducing along any of its symmetries.
All the symmetries generating this algebra become isometries of the
possible metrics (though none of them is a consistent vacuum of the
effective theory).
This means that upon reducing to type IIA along these isometries one
would find a constant dilaton $\phi(x,y) = \phi(x)$ and possibly a
non-trivial 2-form flux.
This implies that the reduced space is a twisted torus; indeed, this
can be explicitly verified.

For this type of reductions, we can identify the 7 moduli
$T_i,i=1,\ldots,7$ of $X_7$ with the type IIA ones of
\cite{Derendinger:2004jn} and compare the respective superpotentials.
In \cite{Derendinger:2004jn}, superpotentials for a type IIA
orientifold of $\mathbb{T}^6/(\mathbb{Z}_2 \times \mathbb{Z}_2)$ with
both physical and geometrical fluxes were constructed from consistent
${\cal N}=1$ truncations of gauged ${\cal N}=4$ supergravity.
Reduction of $X_7$ along an isometry which gives a constant dilaton
reproduces their setup.
We now fix $y^{11}$ as the Killing direction and require that
$\tau_{11 I}^{J} = 0$ in order to have a constant dilaton.

The starting point for connecting with the results of
\cite{Derendinger:2004jn} is to consider our 7-dimensional twisted
toroidal orbifold as a fibration of the form (\ref{metricX7}), where
the base is a twisted 6-dimensional orbifold
$\mathbb{T}^6/(\mathbb{Z}_2 \times \mathbb{Z}_2)$.
The latter is naturally equipped with an SU(3)-structure specified by
$J$ and $\Omega$.
Because of our conventions for the G$_2$ 3-form (\ref{Phi}), the SU(3)
decomposition reads
\begin{equation}
\Phi= e^{-\frac{2}{3}\phi} J \wedge \alpha +  e^{-\phi}\; \hat{\rho},
\label{g2}
\end{equation}
with the SU(3)-structure being defined as
\begin{eqnarray}
J&=& t'_1 dy^5 \wedge dy^6 + t'_2 dy^7 \wedge dy^8 + t'_3 dy^9 \wedge 
dy^{10}, \\
\Omega&=&\rho+i \hat \rho=-\sqrt{\frac{t'_1 t'_2 t'_3}{u_1 u_2 u_3}} 
(u_1 dy^5 + i dy^6)
\wedge (u_2 dy^7 + i dy^8) \wedge (u_3 dy^9 + i dy^{10}).
\end{eqnarray}
These SU(3)-structure forms are normalized as $\Omega \wedge
{\overline \Omega}=-\frac{4 i}{3} J \wedge J \wedge J$, so that the
corresponding G$_2$-structure (\ref{g2}) is also properly normalized.
As usual, $t'_i$ and $u_i$ are the K\"ahler and complex structure
moduli respectively, in terms of which the non-vanishing components of
the metric are
\begin{equation}
g_{55}=t'_1 u_1, \;\;g_{66}=t'_1/u_1,\;\;
g_{77}=t'_2 u_2, \;\;g_{88}=t'_2/u_2,\;\;
g_{99}=t'_3 u_3, \;\;g_{1010}=t'_3/u_3. 
\end{equation}

The superpotential of \cite{Derendinger:2004jn} depends on 7 chiral
multiplets $(T'_A,U'_A,S')$, for $A=1,2,3$, whose real parts are
$(t'_A, u'_A,s')$ with
\begin{equation}
s'=\sqrt{\frac{s}{u_1 u_2 u_3}}, \;\;\;u'_1=\sqrt{\frac{s u_2 u_3}{u_1}},
\;\;\;u'_2=\sqrt{\frac{s u_1 u_3}{u_2}}, \;\;\;u'_3=\sqrt{\frac{s u_1 u_2}{u_3}}.
\end{equation}
Here $s$ is the 4-dimensional dilaton defined as $s\equiv e^{-2\phi_4}
=e^{-2\phi} (t'_1 t'_2 t'_3)$.

A comparison with the G$_2$ form
\begin{eqnarray}
 \Phi&=&(t_1 dy^5 \wedge dy^6 - t_2 dy^7 \wedge dy^8  - t_3 dy^9 \wedge 
dy^{10}) 
 \wedge dy^{11} \\
&&- t_4 dy^6 \wedge dy^7 \wedge dy^9   + t_5 dy^6 \wedge dy^8 \wedge dy^{10}   
+ t_6 dy^5 \wedge dy^7 \wedge dy^{10}   + t_7 dy^5 \wedge dy^8 \wedge 
dy^9\nonumber
\end{eqnarray}
yields the following identification of moduli:
\begin{equation}
t_1=t'_1,\;\; t_2=-t'_2,\;\; t_3=-t'_3, \;\; t_4=u'_1,\;\;t_5=s',\;\;t_6=-u'_3,\;\;
t_7=-u'_2.
\label{ident}
\end{equation}
This obviously extends to the full chiral multiplets.
Turning on the geometric fluxes amounts to replacing $dy^I$ by
$\eta^I$ and hence the identifications remain the same.

The decomposition of fluxes works straightforwardly.
The allowed 4-form fluxes in 11-dimensions
\begin{eqnarray}
&g_{78910},\;\; g_{56910},\;\;g_{5678},& \\
&g_{581011},\;\;g_{671011},\;\;g_{57911},\;\;g_{68911},&
\end{eqnarray}
reduce to 4- and 3-form fluxes in type IIA:
\begin{eqnarray}
&f_{78910},\;\; f_{56910},\;\;f_{5678},& \\
&h_{5810},\;\;h_{6710},\;\;h_{579},\;\;h_{689}.&
\end{eqnarray}
The geometrical fluxes of the form $\tau^{11}_{IJ}$ correspond to
2-form RR fluxes since $\eta^{11}=dy^{11}+\sigma \Rightarrow
d\eta^{11}=f_{(2)}=\frac{1}{2} \tau^{11}_{IJ} \eta^{I} \wedge
\eta^{J}$.
For the twisted 7-torus, these are
\begin{equation}
f_{56},\;\; f_{78},\;\;f_{910}.
\end{equation} 
Finally, the 7-form flux $g_{567891011}$ obviously reduces to a 6-form
flux
\begin{equation}
f_{5678910}.
\end{equation}   

We can now re-write the M-theory superpotential (\ref{Wexpanded}),
using the identifications (\ref{ident})\footnote{ The superfields here
correspond to the primed ones in (\ref{ident}) and are precisely those
that appear in \cite{Derendinger:2004jn}.
}:
\begin{eqnarray}
4{\cal W}_{IIA}&=&f_{5678910}+ 
i (T_1 f_{78910}+ T_2 f_{56910}+ T_3 f_{5678})+ i S h_{579}\nonumber\\
&-&i (U_1 h_{5810}+ U_2 h_{6710}+ U_3 h_{689}) - (T_1 T_2 f_{9 10} + T_2 T_3 f_{56}+
T_3 T_1 f_{78})\\
&+& S( T_1 \tau^6_{79} +   T_2 \tau^8_{95} + T_3 \tau^{10}_{57})
-(T_1 U_1 \tau^6_{8 10} + T_2 U_2 \tau^8_{10 6}+ T_3 U_3 \tau^{10}_{6 8})\nonumber\\
&+& (T_1 U_2 \tau^5_{7 10} + T_1 U_3 \tau^5_{8 9} + T_2 U_1 \tau^7_{10 5} +
T_2 U_3 \tau^7_{9 6}+T_3 U_1 \tau^9_{5 8}+T_3 U_2 \tau^9_{6 
7}).\nonumber
\end{eqnarray}
The above expression contains all the terms of the superpotential of
\cite{Derendinger:2004jn}\footnote{The sign difference in the terms
involving the geometrical fluxes is due to our sign convention in
(\ref{torsion}).} except for the cubic one, since this corresponds to
a massive type IIA reduction, which, as we said earlier, should be
invisible in our approach.
The two superpotentials match, as we have seen, because $\tau_{11
I}^{J} = 0$.

Assuming a similar identification of the type IIA moduli fields as
above, the terms in (\ref{Wexpanded}) involving $\tau_{11 I}^{J}$
would mean that couplings of the form $SU$ and $UU$ appear in the IIA
superpotential.
These are not present in \cite{Derendinger:2004jn} and therefore we should
expect them when the type IIA theory is compactified on an internal 
manifold which is not a twisted torus.
It would be interesting to give a concrete example of this phenomenon,
but as this requires a further analysis of the various quotient group
manifolds presented here we postpone it for future work.

\section{Summary and outlook}

In this paper we studied the Scherk--Schwarz reduction of M-theory in
the presence of fluxes, using the equivalence between Scherk--Schwarz
reductions and compactifications on twisted tori.
The latter allows us to go beyond the realm of exceptional holonomy,
in this case G$_2$, in a concrete and controllable framework.
In particular, the 11-dimensional supergravity action can be
explicitly reduced and we can obtain the 4-dimensional effective
potential for the light modes.
This result is the sum of the expressions (\ref{Einstein}),
(\ref{V4form}) and (\ref{VCS}).
Now, the fact that our models have precisely a G$_2$-structure implies
that the 4-dimensional theory is an ${\cal N}=1$ supergravity and
hence that the potential should be derivable by a ${\cal N}=1$
superpotential.
Indeed, the superpotential (\ref{Wexpanded}) reproduces precisely the
potential we found from the explicit reduction, using also the
K\"ahler potential (\ref{Kaehlerbis}).
Furthermore, the K\"ahler potential and superpotential can be written
in terms of the geometric structure of the 7-manifold, in the compact
form (\ref{Kaehler}) and (\ref{Wform}) respectively.
These are in agreement and extend expressions that have been proposed
previously.

The analysis of the vacua is hindered by the fact that the
coefficients of the superpotential depend on the structure constants
that define the twisting, and the latter satisfy the highly
non-trivial constraints (\ref{const}).
Hence, we first presented some generic features concerning the vacuum
structure of superpotentials of the form (\ref{genW}).
This form is the most general one that can be obtained from M-theory
compactifications keeping only the leading order terms in the
11-dimensional supergravity action.
We found that the matrix that determines the quadratic part has to be
degenerate in order to have supersymmetric Minkowski vacua and that it
has to yield dependence on all moduli in order to have supersymmetric
AdS vacua.
Notice that the latter condition is stronger than the fact that the
superpotential has to depend on all moduli in order to have
supersymmetric AdS solutions.
We were able also to show that, in agreement with the no-go theorem of
\cite{deWit:1987xg,Maldacena:2000mw}, supersymmetric Minkowski/AdS
solutions can be obtained only from twisted 7-tori with G$_2$/weak
G$_2$-holonomy unless sources are introduced or higher order 
derivative corrections are considered.

Subsequently we presented some classes of solutions to the constraints
(\ref{const}).
Despite the wealth of solutions, the best that can be achieved are
supersymmetric and non-supersymmetric Minkowski vacua with at most 3
moduli stabilized.
However, our study revealed several interesting properties of this
type of reductions.
Among others, a geometric way of understanding the mechanism behind
moduli fixing in Scherk--Schwarz compactifications was obtained, using
the flat groups as toy models.
As a by-product, the difference between ordinary Kaluza--Klein
compactification on a twisted torus (which would yield seven massless
moduli for a flat group) and Scherk--Schwarz reduction (which yields
four massless moduli) was elucidated.
The underlying principle is that the expansion of all physical fields
in internal pieces and external moduli fields should respect a certain
symmetry group.
The choice of this group selects the fields that appear in the
effective action and in some cases extra care is required in order to
identify correctly the massless degrees of freedom.

Another point concerns the introduction of fluxes.
Although it is a priori assumed that the background fluxes are
harmonic, in cases where the internal
manifold is a different parametrization of flat space the flux
evaluated at the solution is vanishing.
Again this is a consequence of the no-go theorem.
Also, it can be seen explicitly from our supersymmetry condition
$M_{ij} \tau^j + G_i = 0$, which fixes the axion values in terms of
the 4-form fluxes.
This relation, valid for both Minkowski and AdS vacua, once expressed
using differential forms and the relations between $M_{ij}$ and
$\tau_{IJ}^{K}$, implies the exactness of the allowed 4-form fluxes.
Consequently,  since by assumption the 4-form fluxes $G_i$ are harmonic,
they are identically zero.

A consistency check of our approach is performed by reducing our
results to type IIA theory.
In this way we were able to derive both the generic form of the type
IIA superpotential for a compactification on a manifold with 
SU(3)-structure and also the superpotentials of \cite{Derendinger:2004jn}.
The latter were derived in a complementary bottom-up approach, which
utilized the machinery of ${\cal N}=4$ supergravity and its ${\cal
N}=1$ truncations.
Our top-to-bottom approach hints towards the existence of more quadratic couplings
between the type IIA moduli than those  found in \cite{Derendinger:2004jn},
when a non-trivial dilaton is present and the internal 6-manifolds are
not twisted tori.
A more precise analysis of these geometries is left for future work.

Another interesting open problem in this context is to establish a
dictionary between the effective 4-dimensional approach and the one we
have followed here (including also sources).
It would be worthwhile, for example, to understand the correspondence
between the constraints imposed on the structure constants of the
${\cal N}=1$ gauging by the Jacobi identities and the 10- or
11-dimensional constraints involving the geometrical and physical
fluxes (the analogues of (\ref{const}) and (\ref{IB})).
This will clarify whether the techniques based on gauged supergravity
are able to capture all possible effective supergravities and, vice
versa, the extent to which the ${\cal N}=1$ gauged supergravities
actually admit a 10- or 11-dimensional space--time interpretation.

Understanding the deformations of G$_2$-structures is another urgent
task that will help us obtain a better picture of the landscape of
${\cal N}=1$ M-theory vacua.
As we have already pointed out, the K\"ahler potential (\ref{Kaehler})
is the Hitchin functional on the space of stable 3-forms.
It has been argued that a similar Hitchin functional, which appears as
the K\"ahler potential for type II compactifications on 6-manifolds
with SU(3)-structure, describes the moduli space of SU(3)-structures
\cite{Grana:2005ny}.
It would be extremely interesting to extend the analysis of
\cite{Grana:2005ny} for 7-manifolds with G$_2$-structure.

Finally, it would be important to extend the considerations of this
paper to other concrete examples of manifolds with  G$_2$-structure.
In particular, one can consider freely acting  G$_2$-orbifolds
\cite{Acharya:1998pm} which are not plagued by extra blow-up moduli.
Analyzing the possibilities for moduli stabilization due to 
Scherk--Schwarz deformations and form fluxes in these examples
is left for future work.


\bigskip \bigskip

\noindent
{\bf Acknowledgments}

\medskip

We are grateful and indebted to B.~S.~Acharya for many useful
suggestions and collaboration at the early stages of this work.
We would also like to thank C.~Angelantonj, R.~D'Auria,
J.-P.~Derendinger, S.~Ferrara, D.~Joyce, D.~L\"ust, P.~Manousselis,
M.~Schnabl, K.~Sfetsos, S.~Stieberger, G.~Villadoro, F.~Zwirner and
especially A.~Uranga for enlightening discussions.
The work of N.P.
has been supported by the Swiss National Science Foundation and by the
Commission of the European Communities under contract
MRTN-CT-2004-005104.

\bigskip

\bigskip

\appendix
\section*{Appendix: Dual gauged supergravity algebras from the M-theory 
pseudo-action}

\makeatletter \@addtoreset{equation}{section} \makeatother
\renewcommand{\theequation}{A.\arabic{equation}}

As we have seen in section \ref{secpotential}, the pseudo-action
(\ref{pseudoaction}) can be very useful to describe part of the
degrees of freedom of the 3-form, using the dual 6-form as well as
introducing the dual form-fluxes.
Here we will see, for the general case of Scherk--Schwarz
compactifications of M-theory on ${\mathbb T}^7$, how this formalism
let us derive the effective gauged supergravity algebras in the dual
form of \cite{DallAgata:2005mj,DAuria:2005er}, without the need of
cumbersome group-theoretical arguments.
This means that we can naturally integrate out the massive tensor
degrees of freedom appearing in the standard compactifications.

For this purpose we will continue to follow the convention on the
splitting of the 11-dimensional indices in space--time ones
$\mu,\nu\ldots$ and internal ones $I,J,K,\ldots$.
Also, in order to describe effectively the 4-dimensional degrees of
freedom without introducing massive tensor fields, we have to choose
an appropriate set of connections and curvatures to be plugged in the
pseudo-action (\ref{pseudoaction}).

The main point in this choice is that consistency implies that not all
vectors can be in the ${\bf 21}$ of $sl(7)$, i.e.~described by the
$C_{\mu IJ}$ connection; some of them  (up to seven), those which will
become massive by eating the scalars dual to the tensor fields, should
be described by the dual connection $A_{\mu IJKLM}$.
For this reason one has to split the couples of indices $IJ$ into
those belonging to closed forms (hence related to the 2-cycles ${\cal
C}_2$), and the orthogonal ones. 
We will see the reasons for this choice in the following.
These conditions will turn out to be consistency conditions
required in order to make the formalism work.
Since we do not want tensor fields to appear naked in the effective
theory coming from the pseudo-action, we choose to turn on the
curvatures
\begin{equation}
G_{\mu\nu IJ}, \quad G_{\mu IJK}, \quad G_{IJKL}, \quad \widetilde 
G_{\mu\nu IJKLM}, \quad \widetilde 
G_{\mu IJKLMN}, \quad \widetilde G_{IJKLMNP},
\label{dualcurvM5}
\end{equation}
and the dual connections
\begin{equation}
C_{\mu\nu\rho}, \quad C_{\mu\nu I}, \quad C_{\mu IJ}, \quad
A_{\mu\nu\rho\sigma IJ}, \quad A_{\mu\nu\rho IJK}, \quad A_{\mu\nu
IJKL}, \quad A_{\mu IJKLM},
\label{connM5}
\end{equation}
where for the moment we leave the possibility of having both the standard
vectors $C_{\mu IJ}$ and the dual ones $A_{\mu IJKLM}$.

The variation of the (\ref{pseudoaction}) pseudo-action 
with respect to (\ref{connM5}) gives
\begin{eqnarray}
 &  & \partial_{\sigma} \widetilde G_{IJKLMNP} - \frac72
 \tau_{[IJ}^{Q}\widetilde G_{\sigma KLMNP]Q}+ 70 G_{\sigma 
 [IJK}G_{LMNP]} = 0, \label{ib1}\\[2mm]
 &  & \partial_{[\rho}\widetilde G_{\sigma] IJKLMN}  + 20 G_{[\rho [IJK} 
 G_{\sigma] LMN]} + 15 G_{\rho\sigma [IJ} G_{KLMN]} = 0, \label{ib2}\\[2mm]
 &  & \partial_{[\nu} \widetilde G_{\rho\sigma] IJKLM} + 20 G_{[\nu 
 [IJK}G_{\rho\sigma] LM]} =0, \label{ib0}\\[2mm]
 &  & \tau_{[IJ}^{P} G_{KLM]P} = 0,
 \label{ib3}\\[2mm]
 &  & \partial_{\mu} G_{IJKL} - 6 \tau_{[IJ}^{P} G_{\mu KL]P}=0,\label{ib4}\\[2mm]
 &  & \partial_{[\rho}G_{\sigma] IJK} -\frac32 \tau_{[IJ}^{P} G_{\rho\sigma 
 K]P}=0, \label{ib5}\\[2mm]
 &  & \partial_{[\nu}G_{\rho\sigma] IJ} = 0. \label{ib6}
\end{eqnarray}
We can start solving them from the last one, inserting each time
the solution in the previous equation.
In this way, we find that the curvatures should be defined by
\begin{eqnarray}
G_{\mu\nu IJ} &=& 2 \partial_{[\mu} C_{\nu] IJ}, \label{sol1}\\[2mm]
G_{\mu IJK} &=& \partial_{\mu} C_{IJK} +3 \tau_{[IJ}^{P} C_{\mu 
 K]P}, \label{sol2}\\[2mm]
G_{IJKL} &=& g_{IJKL} + 6 \tau_{[IJ}^{P}C_{KL]P}, \label{sol3}\\[2mm]
\tau_{[IJ}^{P} g_{KLMN]} &=&0,
 \label{sol4}\\[2mm]
\widetilde G_{\mu\nu IJKLM} &=& 2 \partial_{[\mu} A_{\nu] IJKLM} - 60  
 C_{[\mu [IJ} \tau_{KL}^{P} C_{\nu]M]P} - 40 C_{[\mu [IJ} 
 \partial_{\nu]} C_{KLM]}, \label{sol0} \\[2mm]
\widetilde G_{\mu IJKLMN} &=& \partial_{\mu} A_{IJKLMN} + 15 \, 
 \tau_{[IJ}^{P} A_{\mu KLMN]P}-20 
C_{[IJK}\partial_{\mu} C_{LMN]} \nonumber \\[2mm]
&-&30 \,C_{\mu [IJ}\left(g_{KLMN]} + 6 
\tau_{KL}^{P} C_{MN]P}\right),  \label{sol5}\\[2mm]
\widetilde G_{IJKLMNP} &=& \tilde g_{IJKLMNP}+ 21
 \tau_{[IJ}^{Q}A_{KLMNP]Q} - 70  C_{[IJK} \left(g_{LMNP]} + 3
 \tau_{LM}^{Q} C_{NP]Q}\right),\label{sol6}
\end{eqnarray}
and are subject to the constraint
\begin{eqnarray}
\tau \left( C_{[\mu [IJ} \partial_{\nu} C_{\rho] KL]}\right) & = & 0.
\label{con1dual}
\end{eqnarray}
This is a very strong constraint on the allowed vector fields.
It appears because we are varying with respect to an unconstrained set
of $C_{\mu IJ}$.
On the other hand, we know that only a subset of them should be
dualized.
This set can be chosen precisely in a way which gets rid of the
constraint (\ref{con1dual}).
By splitting $C_{\mu IJ} = \tau_{IJ}^{K} C_{\mu K} + C_{\mu IJ}^{0}$
\cite{DAuria:2005er}, we see that we do not get the constraint
(\ref{con1dual}) when varying the pseudo-action (\ref{pseudoaction})
with respect to $C_{\mu K}$, as this term appears in a total
derivative on the internal compact space.
This is precisely what is required by the dualization procedure.
The (up to 7) vectors $C_{\mu K}$ have to be dualized, while the
$C_{\mu IJ}^{0}$ components appear in the final action.

From these solutions we can see two important consequences for the
effective theory.
The first one is that there are no space--time tensor fields left.
The standard degrees of freedom of the 7 tensor fields $C_{\mu\nu I}$
are replaced by the dual 7 scalars $A^{P} \equiv \epsilon^{PIJKLMN}
A_{IJKLMN}$.
It is also explicitly made clear here that these are in the conjugate
representation of $SL(7)$ with respect to the original ones (if the
tensors are in the {\bf 7}, the dual scalars are in the $\overline
{\bf 7}$).
Also, the 21 vector fields are split in two sets (as explained above).
A number $n \leq 7$ of them are described by the dual vectors coming
from the 6-form potential $\tilde A_\mu^{Q} = \epsilon^{IJKLMNP}
\tau_{IJ}^{Q} A_{\mu KLMNP}$, whereas the remaining $21 -n$ are
described by the original 3-form $C_{\mu IJ}$.
This is the only way to close consistently the above Bianchi
identities.
If all the independent components of the vectors $C_{\mu IJ}$ are
expanded on the 2-cycles, then one can solve directly the Bianchi
identities (\ref{ib1})--(\ref{ib6}) without the need for introducing
the dual potentials.

The second important aspect is given by the relation (\ref{sol4}).
Since the constant fluxes of the 4-form fields act as coupling
constants in the effective theory, this relation implies a change in
the structure constants of this theory.
At this stage we can plug these solutions back in the original action
and therefore also complete the effective theory gauged supergravity
algebra.
As in \cite{DallAgata:2005ff,Hull:2005hk} this can be obtained by
looking at the commutators of the gauge transformations of the vector
fields as they are in a faithful representation of the gauge group.
These transformations read
\begin{equation}
\begin{array}{rcl}
\delta G_{\mu}^{I} &=& \partial_{\mu} \omega^{I}-\tau_{JK}^{I}
\omega^{J} G_{\mu}^{K}, \\[3mm] \delta C_{\mu IJ} &=& \partial_\mu
\Sigma_{IJ} + 3 G_\mu^P \tau_{[PI}^S \Sigma_{J]S} - 2 \omega^Q
\tau_{Q[I}^S C_{\mu J] S} - \omega^R g_{RIJQ} G_\mu^Q, \\[3mm] \delta
A_{\mu IJKLM} &=& \partial_{\mu} \Theta_{IJKLM} -6 G_\mu^P
\tau_{[PI}^S \Theta_{JKLM]S} +5 \omega^Q \tau_{Q[I}^S A_{\mu JKLM] S}
\\[2mm] &+& \omega^R g_{RIJKLMQ} G_\mu^Q - 10 \Sigma_{[IJ}
\partial_{\mu} C_{KLM]}-10
C_{\mu [IJ} \omega^{Q} g_{KLM]Q}\\[2mm] 
&-& 12 \Sigma_{[IJ}\left(2
g_{KLMP]} + 3 \tau_{MP}^S C_{KLS}\right) G_{\mu}^{P} - 10 C_{[IJK} \omega^{P} g_{LM]NP}
G_{\mu}^{P}.
\end{array}
\label{vecttransf}
\end{equation}
The algebra can thus be obtained by assigning the generators 
\begin{equation}
T_{A} = \{Z_{I}, W^{IJ}, W_{IJ}\},
\label{generators}
\end{equation}
matching the parameters $\{\omega^{I}, \Sigma_{IJ}, \epsilon^{IJKLMNP}
\Theta_{KLMNP}\}$ respectively.
The commutators of these generators give 
\begin{equation}
\begin{array}{rcl}
[Z_I, Z_J] &=& \tau_{IJ}^{K} Z_K + g_{IJKL} W^{KL} + 
g_{IJKLMNP} \epsilon^{KLMNPQR}W_{QR}, 
\\[2mm]
[Z_I, W^{JK}] &=& 2 \tau_{IM}^{[J}W^{K]M} -10 \epsilon^{JKLMNPQ} g_{ILMN} 
W_{PQ},\\[2mm]
[Z_{I}, W_{JK}] &=& 2 \tau_{I[J}^{P} W_{K]P} = \tau_{JK}^{P}W_{PI}, \\[2mm]
[W^{IJ}, W^{KL}] &=& 30\, W_{PQ} 
\epsilon^{MNPQIJ[K} \tau_{MN}^{L]},\\[2mm]
[W^{IJ}, W_{KL}] &=& [W_{IJ}, W_{KL}]= 0,
\end{array}
\label{algebra}
\end{equation}
and they constitute the gauge algebra in the dual formulation, which
was also obtained in a similar form using group-theoretical
arguments in \cite{DAuria:2005er}.
The closure of the algebra can be obtained by recalling the split in the 
vectors, which removes some problematic terms in the commutators of 
the gauge transformations of $C_{\mu IJ} = C_{\mu IJ}^{0}$.
Also, one has to use some remarkable relations,
first found in \cite{DAuria:2005er},
among the algebra 
generators : 
$\tau_{[IJ}^{P}W_{K]P} = 0$ and $10\, \epsilon^{ABCDIJK} g_{ABCD}W_{JK} = 
-2 \tau_{JK}^{I} W^{JK}$.


\providecommand{\href}[2]{#2}

\end{document}